# Chordwise micro-jet placement reveals a trade-off between mean hydrodynamic performance and unsteady loading under cloud cavitation

Amirmehran Mahdavi[1], Soheil Motezakkeri[1], Yasin Ebrahimi[1], Ali Alavi[2]

1. Department of Mechanical Engineering, Hakim Sabzevari University, Sabzevar, Iran
2. High Performance Computing Laboratory, Department of Mechanical Engineering, Ferdowsi University of Mashhad, Mashhad, Iran

*Corresponding author: am.me.mahdavi@gmail.com*

*am.mahdavi@hsu.ac.ir*

## Abstract

Prescribed tangential micro-jet injection can modify cloud-cavitation dynamics by redistributing near-wall momentum, but the injection location that benefits mean hydrodynamic performance need not be the location that minimizes unsteady loading. This study systematically compares five chordwise locations (x/c = 0.15, 0.30, 0.45, 0.60, and 0.70) over a Clark-Y hydrofoil at Re = $7 \times 10^5$ and $\sigma = 0.8$ under an identical prescribed tangential-jet condition. Transient large-eddy simulation coupled with a Volume-of-Fluid formulation and the Schnerr-Sauer cavitation model is used in a two-dimensional parametric framework, supplemented by one representative three-dimensional calculation to illustrate spanwise cavity deformation. The analysis combines vapor topology, normalized turbulent kinetic energy, velocity and pressure fields, cycle-averaged surface pressure, hydrodynamic forces, record-based force fluctuations, force-response spectra, and cavity-thickness histories. Injection at x/c = 0.15 yields the highest cycle-averaged lift-to-drag ratio among the investigated cases: the mean drag coefficient decreases from 0.137 to 0.107 (21.9%) and $C_L/C_D$ increases from 5.693 to 6.261 (approximately 10.0%), while $C_L$ decreases from 0.78 to 0.67 because of the accompanying suction-side pressure redistribution. In contrast, x/c = 0.60 gives the lowest recorded force-fluctuation RMS, with $C_{L,RMS} \approx 0.112$ and $C_{D,RMS} \approx 0.0165$. The flow-field results indicate that forward injection acts earlier on the developing cavity and near-wall momentum deficit, whereas downstream injection modifies a later stage of cavity evolution. The preferred chordwise location is therefore objective-dependent: the location that maximizes cycle-averaged hydrodynamic efficiency does not coincide with the location that minimizes recorded unsteady loading under the present conditions.

### Nomenclature

| ***Roman symbols*** | | ***Greek symbols*** | |
|---|---|---|---|
| $c$ | Hydrofoil chord, m | $\alpha$ | Angle of attack, ° |
| $C_{cond}$ | Schnerr-Sauer condensation coefficient | $\alpha_l$ | Liquid volume fraction |
| $C_D$ | Drag coefficient | $\alpha_v$ | Vapor volume fraction |
| $C'_D$ | Drag-coefficient fluctuation about record mean | $\Delta$ | LES grid-filter scale, m |
| $C_{D,RMS}$ | RMS drag-coefficient fluctuation | $\Delta t_s$ | Force-history resampling interval, s |
| $C_{evap}$ | Schnerr-Sauer evaporation coefficient | $\delta_{ij}$ | Kronecker delta |
| $C_L$ | Lift coefficient | $\kappa$ | von Karman constant |
| $C'_L$ | Lift-coefficient fluctuation about record mean | $\mu_l$ | Liquid dynamic viscosity, Pa s |
| $C_{L,RMS}$ | RMS lift-coefficient fluctuation | $\mu_m$ | Mixture dynamic viscosity, Pa s |
| $C_\mu$ | Jet momentum coefficient | $\mu_t$ | SGS eddy viscosity, Pa s |
| $C_p$ | Pressure coefficient | $\mu_v$ | Vapor dynamic viscosity, Pa s |
| $C_s$ | Smagorinsky coefficient (0.1) | $\rho_l$ | Liquid density, kg m$^{-3}$ |
| $d$ | Nearest-wall distance in LES mixing-length model, m | $\rho_m$ | Mixture density, kg m$^{-3}$ |
| $D$ | Drag force per unit span, N m$^{-1}$ | $\rho_v$ | Vapor density, kg m$^{-3}$ |
| $f_f$ | Dominant force-response frequency, Hz | $\sigma$ | Cavitation number |
| $k$ | Turbulent kinetic energy, m$^2$ s$^{-2}$ | $\tau_{ij}^{SGS}$ | Subgrid-scale stress tensor, Pa |
| $k^*$ | Globally normalized TKE | ***Dimensionless ratios and coordinates*** | |
| $k_{max,global}$ | Global maximum TKE used for normalization, m$^2$ s$^{-2}$ | $C_L/C_D$ | Cycle-averaged lift-to-drag ratio |
| $L$ | Lift force per unit span, N m$^{-1}$ | $d/c$ | Non-dimensional cavity thickness |
| $L_s$ | SGS mixing length, m | $t/T$ | Case-specific normalized cavitation-cycle coordinate |
| $\dot{m}_{lv}$ | Liquid-to-vapor mass-transfer source, kg m$^{-3}$ s$^{-1}$ | $U_{jet}/U_\infty$ | Jet-to-free-stream velocity ratio |
| $\dot{m}_{vl}$ | Vapor-to-liquid mass-transfer source, kg m$^{-3}$ s$^{-1}$ | $x/c$ | Normalized chordwise location/station |
| $n_b$ | Bubble number density, m$^{-3}$ | ***Abbreviations*** | |
| $p$ | Local pressure, Pa | 2D | Two-dimensional |
| $p_v$ | Saturation/vaporization pressure, Pa | 3D | Three-dimensional |
| $p_\infty$ | Reference pressure for σ and Cp, Pa | CFD | Computational fluid dynamics |
| $R_B$ | Representative bubble radius, m | LES | Large Eddy Simulation |
| $Re$ | Chord Reynolds number | PISO | Pressure-Implicit with Splitting of Operators |
| $\lvert\tilde{S}\rvert$ | Resolved strain-rate magnitude, s$^{-1}$ | PSD | Power spectral density |
| $\tilde{S}_{ij}$ | Favre-filtered resolved strain-rate tensor, s$^{-1}$ | RMS | Root mean square |
| $St_f$ | Force-response Strouhal number | SGS | Subgrid scale |
| $T$ | Case-specific cavitation-cycle duration, s | TKE | Turbulent kinetic energy |
| $T_r$ | Analyzed force-record duration, s | VOF | Volume of Fluid |
| $t$ | Physical time, s | ***Subscripts and superscripts*** | |
| $t_0$ | Initial time of analyzed record, s | *B; cond; evap* | Bubble; condensation; evaporation |
| $u$ | Mixture velocity, m s$^{-1}$ | *D; L* | Drag-related; lift-related |
| $\tilde{u}_i$ | Favre-filtered velocity component, m s$^{-1}$ | *f* | Force-response quantity |
| $U_\infty$ | Free-stream velocity, m s$^{-1}$ | *global; max* | Global value; maximum value |
| $U_{jet}$ | Prescribed jet velocity, m s$^{-1}$ | *i, j, k* | Cartesian/tensor indices |
| $w$ | Micro-jet slot width, m | *jet* | Injection-jet quantity |
| $x$ | Chordwise/spatial coordinate, m | *l; v; m* | Liquid; vapor; mixture |
| $x_i$ | Spatial coordinate in direction i, m | *lv; vl* | Liquid-to-vapor; vapor-to-liquid transfer |
| $x_j$ | Spatial coordinate in direction j, m | *r* | Analyzed force-history record |
| $y^+$ | Dimensionless wall coordinate | *RMS* | Root-mean-square fluctuation |
| | | *s* | Sampling/resampling; SGS mixing-length subscript |
| | | *SGS* | Subgrid-scale quantity |
| | | $\infty$ | Free-stream/reference-flow quantity |
| | | *overbar* | Spatial filter; record mean where stated |
| | | *tilde* | Favre-filtered quantity |
| | | *prime (′)* | Fluctuating component about record mean |
| | | *asterisk (*)* | Normalized quantity |
| | | *superscript T* | Transpose |
| | | *superscript +* | Dimensionless wall-coordinate notation |

## 1. Introduction

Cavitation remains a major hydrodynamic limitation in marine and hydraulic systems because its consequences extend well beyond the local formation of vapor. On hydrofoils, propulsors, pump blades, and other lifting surfaces, attached sheet cavities and their subsequent breakup can substantially modify the surface-pressure distribution, hydrodynamic loads, and wake dynamics. Cloud cavitation is particularly detrimental because the attached cavity undergoes pronounced unsteadiness and periodically sheds large vapor structures into the downstream flow, leading to strong load fluctuations and contributing to vibration, noise, performance deterioration, and potential material damage [1,2]. These effects can compromise both the efficiency and operational reliability of hydraulic and marine components. Consequently, effective cavitation control requires not only limiting the extent of vapor formation but also understanding and modifying the physical mechanisms responsible for cavity instability and large-scale cloud shedding.

The unsteady dynamics of sheet and cloud cavitation result from strongly coupled interactions among the attached cavity, near-wall liquid motion, pressure recovery, and surrounding vortical structures. In many partial- and cloud-cavitation regimes, the adverse pressure gradient near the cavity closure drives an upstream-moving re-entrant liquid layer beneath the attached vapor sheet.

As this re-entrant flow propagates toward the leading edge, it can destabilize and eventually sever the attached cavity, producing a detached vapor cloud that is convected downstream while a new sheet cavity begins to develop [1]. However, re-entrant flow is not the only mechanism capable of sustaining cavitation instability. Compressible simulations of a Clark-Y hydrofoil have shown that pressure disturbances generated by local cloud collapse can interact with the remaining vapor structures and contribute to subsequent cavity destabilization [3]. Three-dimensional investigations further demonstrate a strong two-way coupling between cavitation and vortical dynamics, in which cavity development modifies vortex production, stretching, and deformation, while the evolving vortical field influences cavity growth, breakup, and shedding [4,5]. Accordingly, cavitation-control performance should be evaluated not only through changes in mean cavity extent but also through its influence on re-entrant-flow development, cavity topology, unsteady hydrodynamic loading, and the associated turbulent flow structures.

Numerical simulation has become an important tool for investigating cloud cavitation because it provides access to transient flow quantities and cavity structures that are difficult to characterize comprehensively through measurements alone. In particular, Large Eddy Simulation (LES) is well suited to strongly unsteady cavitating flows because it resolves the energetic turbulent structures that interact with the vapor phase while modeling the effects of unresolved subgrid-scale motions. Previous LES studies have reproduced important features of hydrofoil cavitation, including periodic cavity shedding, transient cavitation–vortex interactions, and the evolution of complex vapor structures [5,6]. For the Clark-Y hydrofoil specifically, verification and validation studies have demonstrated that LES can provide useful predictions of attached and unsteady cavitating flows when adequate spatial and temporal resolution and appropriate cavitation modeling are employed [7]. Nevertheless, the predictive capability of LES is not determined by the turbulence formulation alone; it also depends on mesh resolution, time-step selection, phase-change modeling, interface treatment, and validation against reliable experimental data. These considerations make LES a suitable framework for examining the transient flow mechanisms associated with hydrofoil cavitation control, provided that its numerical limitations are explicitly recognized.

A broad range of flow-control strategies has therefore been developed to mitigate hydrofoil cavitation, which can generally be grouped into passive and active approaches according to the manner in which the surrounding flow is modified. Passive techniques alter the local hydrodynamic environment without continuous external actuation, for example through surface obstacles, grooves, vortex-generating elements, porous treatments, or other geometric modifications that influence pressure recovery, boundary-layer development, and cavity closure [8,9]. Such methods offer the practical advantage of avoiding an external power supply, although their effectiveness is strongly dependent on geometry and operating condition and may involve hydrodynamic penalties outside the targeted cavitating regime. Active approaches, in contrast, deliberately introduce or remove momentum through mechanisms such as water injection, suction, or ventilation, providing greater control authority over cavity development and shedding [8,10,11]. Water injection is particularly attractive because the injected liquid can interact directly with the near-wall flow and the cavity interface, modifying the momentum balance associated with re-entrant-flow development and cavity breakup. However, its effectiveness depends sensitively on injection parameters such as location, direction, and momentum input, motivating a more detailed examination of how localized jet forcing influences cloud-cavitation dynamics.

Jet-based cavitation control has consequently received increasing attention, and previous studies have established that the response depends strongly on where and how the injected momentum interacts with the developing cavity. For a Clark-Y hydrofoil, Yan et al. [12] examined jet location, velocity ratio, and injection direction and showed that these parameters substantially influence both cavitation suppression and hydrodynamic performance. Wang et al. [13] likewise identified injection location as a dominant parameter in a broader assessment of location, angle, and velocity. Gu et al. [14] demonstrated that tangential microjet placement governs the interaction with the re-entrant flow, although favorable locations in their configuration were concentrated around the mid-to-rear chord, while Li et al. [15] further emphasized that intervention position changes the local coupling among injected momentum, the cavity interface, and the near-wall liquid layer. These studies establish chordwise placement as a recognized design variable rather than an unexplored parameter. They also show, however, that favorable locations and control responses remain strongly configuration-dependent. A useful remaining question is therefore how the physical response evolves when a prescribed tangential jet is displaced systematically from the forward to the rear chord under otherwise identical conditions, and whether the location favored by a mean hydrodynamic metric is also the location favored by an unsteady-load metric. A unified comparison linking chordwise position to cavity evolution, near-wall momentum redistribution, turbulent structures, surface-pressure response, mean lift and drag, force fluctuations, and force-response characteristics can help resolve this distinction within a fixed cavitating regime.

Accordingly, the present study investigates five prescribed tangential micro-jet locations spanning $x/c$ = 0.15-0.70 over a Clark-Y hydrofoil under one fixed cavitating operating condition. The two-dimensional simulations are used as a consistent parametric screening framework in which the non-positional jet parameters are held fixed, while a representative three-dimensional calculation is examined separately to illustrate spanwise cavity features that the planar formulation cannot reproduce. The analysis combines transient vapor topology, near-wall velocity and normalized turbulent-kinetic-energy fields, cycle-averaged surface pressure, lift and drag characteristics, cavity-thickness evolution, and record-based force-fluctuation and spectral measures. The study is organized around two related but distinct questions: (i) which chordwise location provides the most favorable cycle-averaged hydrodynamic force balance among the tested cases, and (ii) whether that same location also minimizes the recorded unsteady hydrodynamic loading. By connecting these global measures to the streamwise stage at which momentum is introduced relative to cavity development, the analysis seeks to explain why forward, intermediate, and rearward injection can favor different performance objectives rather than to propose a universal optimum location.

## 2. Governing equations

The unsteady cavitating flow is described using a homogeneous mixture formulation in which the liquid and vapor phases share a common pressure and velocity field. The phase distribution is represented through the local volume fraction within the Volume-of-Fluid (VOF) framework, while the interphase mass transfer associated with evaporation and condensation is subsequently modeled using the Schnerr–Sauer cavitation model. The VOF approach provides a conservative framework for representing the spatial distribution and evolution of the two phases [16] and has been extensively applied to transient cavitating flows over hydrofoils in conjunction with turbulence-resolving approaches such as LES [6]. The governing formulation adopted here is presented in terms of the mixture conservation equations, followed by the vapor-volume-fraction transport equation, the cavitation mass-transfer model, and the LES closure.

### 2.1. Homogeneous mixture formulation

The liquid water and water vapor are treated as two incompressible phases with constant phase properties, whereas the local properties of the mixture vary according to the phase volume fractions. Within the homogeneous-mixture assumption, relative slip between the liquid and vapor phases is neglected and a single velocity field, $\mathbf{u}$, is solved throughout the computational domain. The conservation of mixture mass is expressed as

$$\frac{\partial \rho_m}{\partial t} + \nabla \cdot (\rho_m \mathbf{u}) = 0 \qquad (1)$$

where $\rho_m$ denotes the local mixture density. The corresponding momentum equation is written as

$$\frac{\partial (\rho_m \mathbf{u})}{\partial t} + \nabla \cdot (\rho_m \mathbf{u} \otimes \mathbf{u}) = -\nabla p + \nabla \cdot [\mu_m (\nabla \mathbf{u} + \nabla \mathbf{u}^T)] \qquad (2)$$

where $p$ is the pressure and $\mu_m$ is the local dynamic viscosity of the mixture. The additional stresses arising from the LES filtering operation are introduced separately in the turbulence-model formulation below.

The mixture density and dynamic viscosity are determined by volume-fraction weighting of the liquid and vapor properties,

$$\rho_m = \alpha_l \rho_l + \alpha_v \rho_v \qquad (3)$$

$$\mu_m = \alpha_l \mu_l + \alpha_v \mu_v \qquad (4)$$

subject to

$$\alpha_l + \alpha_v = 1 \qquad (5)$$

where $\alpha_l$ and $\alpha_v$ are the liquid and vapor volume fractions, respectively, and $\rho_l$, $\rho_v$, $\mu_l$, and $\mu_v$ denote the corresponding phase densities and dynamic viscosities. This single-fluid formulation is consistent with the VOF-based treatment commonly employed for numerical simulations of cavitating hydrofoil flows [6,16,17].

### 2.2. VOF formulation and Schnerr–Sauer cavitation model

The temporal evolution of the liquid–vapor distribution is described through the transport of the vapor volume fraction, $\alpha_v$, within the VOF framework. For the homogeneous mixture considered here, the conservative vapor-phase transport equation can be written as

$$\frac{\partial(\alpha_v \rho_v)}{\partial t} + \nabla \cdot (\alpha_v \rho_v \mathbf{u}) = \dot{m}_{lv} - \dot{m}_{vl} \qquad (6)$$

where $\dot{m}_{lv}$ and $\dot{m}_{vl}$ denote the volumetric mass-transfer rates from liquid to vapor and from vapor to liquid, respectively. Since the vapor density is treated as constant, variations in $\alpha_v$ directly describe the local evolution of the vapor phase, while the liquid volume fraction is obtained from $\alpha_l = 1 - \alpha_v$. This transport-equation formulation has been extensively employed for numerical prediction of unsteady hydrofoil cavitation [6,16–18].

The interphase mass transfer is modeled using the Schnerr–Sauer cavitation model [19]. This model relates evaporation and condensation to the dynamics of a population of spherical vapor nuclei and is derived from a simplified form of the Rayleigh–Plesset equation. After neglecting the second-order bubble-acceleration term, viscous contribution, and surface-tension contribution, the bubble-growth or collapse rate is approximated as [19]

$$\frac{dR_B}{dt} = \begin{cases} \sqrt{\frac{2}{3}\frac{p_v - p}{\rho_l}}, & p < p_v, \\ -\sqrt{\frac{2}{3}\frac{p - p_v}{\rho_l}}, & p > p_v. \end{cases} \qquad (7)$$

where $R_B$ is the representative bubble radius, $p$ is the local pressure, and $p_v$ is the saturation vapor pressure of the liquid at the operating temperature. The same bubble-dynamics basis is used in the Fluent implementation of the Schnerr–Sauer model.

Assuming a constant bubble number density $n_b$, the vapor volume fraction is related to the representative bubble radius through

$$\alpha_v = \frac{\frac{4}{3}\pi n_b R_B^3}{1 + \frac{4}{3}\pi n_b R_B^3} \qquad (8)$$

which yields

$$R_B = \left[\frac{3}{4\pi n_b}\frac{\alpha_v}{1-\alpha_v}\right]^{1/3} \qquad (9)$$

Accordingly, the evaporation mass-transfer rate for $p < p_v$ is expressed as

$$\dot{m}_{lv} = C_{\mathrm{evap}}\frac{\rho_l \rho_v}{\rho_m}\alpha_v(1-\alpha_v)\frac{3}{R_B}\sqrt{\frac{2}{3}\frac{p_v - p}{\rho_l}}, \qquad p < p_v \qquad (10)$$

whereas the condensation rate for $p > p_v$ is given by

$$\dot{m}_{vl} = C_{\mathrm{cond}}\frac{\rho_l \rho_v}{\rho_m}\alpha_v(1-\alpha_v)\frac{3}{R_B}\sqrt{\frac{2}{3}\frac{p - p_v}{\rho_l}}, \qquad p > p_v \qquad (11)$$

Here, $C_{\mathrm{evap}}$ and $C_{\mathrm{cond}}$ are the evaporation and condensation coefficients, respectively. Through the dependence on $\alpha_v(1-\alpha_v)$, $R_B$, and the local pressure difference relative to $p_v$, the Schnerr–Sauer formulation couples the macroscopic vapor transport to an idealized description of bubble growth and collapse [19]. Transport-equation cavitation models based on related pressure-driven mass-transfer concepts have been widely assessed in hydraulic and marine applications [20]. For hydrofoil flows, Li et al. [18] employed the Schnerr–Sauer model in simulations of unsteady sheet cavitation, while the more recent verification and validation study of Deng et al. [21] showed that the Schnerr–Sauer model can reproduce the principal cavity-evolution patterns around a Clark-Y hydrofoil, although the quantitative prediction remains dependent on the selected cavitation and turbulence formulations.

In the present ANSYS Fluent implementation, the vaporization pressure was specified as 60,000 Pa and the bubble number density as $1 \times 10^{11}$ $\mathrm{m}^{-3}$. These are the model inputs used for all baseline and controlled calculations.

### 2.3. Large Eddy Simulation formulation

Large Eddy Simulation (LES) is employed to represent the strongly unsteady turbulent structures associated with cavity development, breakup, and shedding. In LES, the large, energy-containing turbulent motions are directly resolved, whereas the effects of motions smaller than the local filter scale are represented through a subgrid-scale (SGS) model [22,23]. This scale-separation concept is particularly relevant to cavitating hydrofoil flows, where large-scale vortical structures interact strongly with the vapor phase and contribute to the temporal evolution of sheet and cloud cavities [24–31].

For the variable-density liquid–vapor mixture, density-weighted (Favre) filtering is introduced according to

$$\tilde{u}_i = \frac{\overline{\rho_m u_i}}{\overline{\rho}_m} \qquad (12)$$

where the overbar denotes spatial filtering and the tilde represents Favre-filtered quantities. Application of this filtering operation to the mixture momentum equation introduces an additional SGS stress tensor, defined as

$$\tau_{ij}^{SGS} = \overline{\rho_m u_i u_j} - \overline{\rho}_m \tilde{u}_i \tilde{u}_j \qquad (13)$$

Accordingly, the filtered momentum equation can be expressed as

$$\frac{\partial(\overline{\rho}_m \tilde{u}_i)}{\partial t} + \frac{\partial(\overline{\rho}_m \tilde{u}_i \tilde{u}_j)}{\partial x_j} = -\frac{\partial \overline{p}}{\partial x_i} + \frac{\partial}{\partial x_j}\left[\overline{\mu}_m \left(\frac{\partial \tilde{u}_i}{\partial x_j} + \frac{\partial \tilde{u}_j}{\partial x_i}\right)\right] - \frac{\partial \tau_{ij}^{SGS}}{\partial x_j} \qquad (14)$$

The unresolved SGS stresses are modeled using the Smagorinsky–Lilly eddy-viscosity formulation [22,23]. The deviatoric component of the SGS stress tensor is related to the resolved strain-rate tensor through

$$\tau_{ij}^{SGS} - \frac{1}{3}\tau_{kk}^{SGS}\delta_{ij} = -2\mu_t \left(\tilde{S}_{ij} - \frac{1}{3}\tilde{S}_{kk}\delta_{ij}\right) \qquad (15)$$

where $\delta_{ij}$ is the Kronecker delta and $\mu_t$ denotes the SGS turbulent viscosity. The resolved strain-rate tensor is defined as

$$\tilde{S}_{ij} = \frac{1}{2}\left(\frac{\partial \tilde{u}_i}{\partial x_j} + \frac{\partial \tilde{u}_j}{\partial x_i}\right) \qquad (16)$$

In the Smagorinsky–Lilly model, the SGS turbulent viscosity is evaluated from

$$\mu_t = \overline{\rho}_m L_s^2 |\tilde{S}| \qquad (17)$$

with

$$|\tilde{S}| = \sqrt{2\tilde{S}_{ij}\tilde{S}_{ij}} \qquad (18)$$

where $L_s$ represents the SGS mixing length. For the standard Smagorinsky–Lilly formulation, this length scale is determined from

$$L_s = \min(\kappa d,\ C_s \Delta) \qquad (19)$$

where $\kappa$ is the von Kármán constant, $d$ is the distance from the nearest wall, $C_s$ is the Smagorinsky coefficient, and $\Delta$ denotes the local grid-filter scale. This formulation limits the SGS length scale in the near-wall region while relating the modeled turbulent viscosity in the outer flow to the local computational resolution. In the present simulations, the standard Smagorinsky–Lilly SGS model was employed using the default model constants provided by ANSYS Fluent. Accordingly, the Smagorinsky constant was set to $C_s = 0.1$, with no additional modification or dynamic adjustment of the SGS coefficient.

LES has been successfully applied to a wide range of cavitating-flow configurations. Wang and Ostoja-Starzewski [24] demonstrated its capability in resolving sheet/cloud cavitation over a NACA0015 hydrofoil, while Ji et al. [25] employed three-dimensional LES to investigate cavity shedding and vorticity evolution around a twisted hydrofoil. Subsequent studies extended LES to attached cavitation and wake dynamics over Clark-Y configurations [26,27] and to three-dimensional sheet–cloud cavitation on twisted hydrofoils [28]. More recent investigations have further used LES to examine the relationship between large-scale turbulent vortices and unsteady cavitation [29], as well as cavitation around dynamically moving hydrofoils [30,31]. Collectively, these studies demonstrate that LES is particularly useful for investigating transient cavitation–turbulence interactions, although its predictive quality remains strongly dependent on spatial resolution, temporal resolution, near-wall treatment, and the selected SGS formulation.

## 3. Numerical methodology

The numerical methodology is organized to preserve a consistent operating condition across the baseline and all controlled cases. The principal parametric study is two-dimensional and isolates the effect of injection location; numerical verification, baseline validation, and a complementary three-dimensional calculation are reported separately so that their roles are not conflated with the chordwise comparison.

### 3.1. Hydrofoil geometry and computational domain

The reference geometry is a Clark-Y hydrofoil with chord length $c$ = 70 mm and a fixed angle of attack $\alpha$ = 8°. This configuration is consistent with established Clark-Y cavitation benchmarks at $Re = 7 \times 10^5$ and provides a well-documented basis for comparison with previous numerical and experimental studies [7,32]. The uncontrolled configuration is denoted Case 1. Five controlled cases place a narrow tangential injection slot on the suction surface at $x/c$ = 0.15, 0.30, 0.45, 0.60, and 0.70 (Cases 2-6, respectively). Fig 1 summarizes the hydrofoil geometry and the chordwise locations examined in the parametric study.

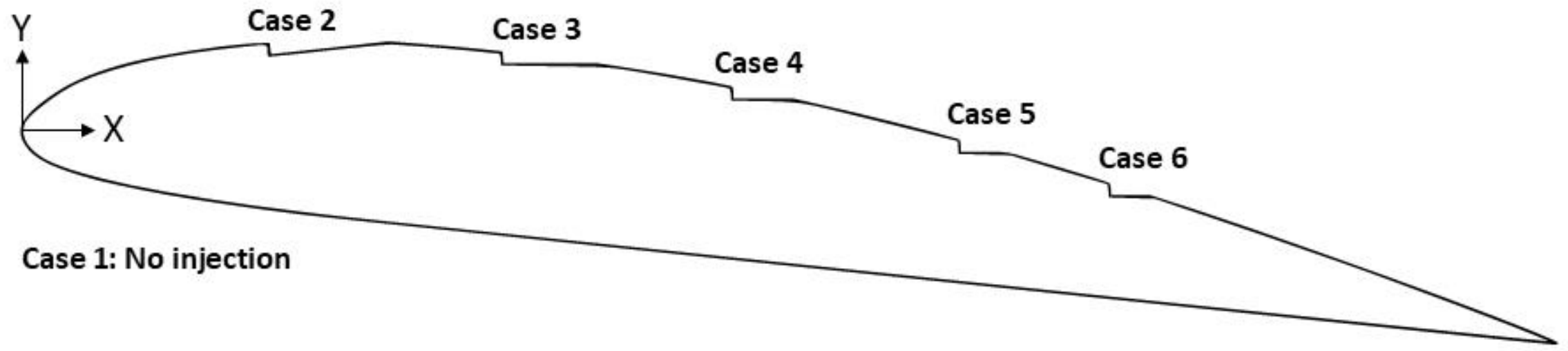

Fig. 1. Clark-Y hydrofoil geometry and chordwise locations of the prescribed tangential micro-jet injection.

The two-dimensional computational domain and its principal boundary types are shown in Fig. 2. A velocity inlet is imposed upstream and a pressure outlet downstream. The upper and lower far-field boundaries are treated as symmetry boundaries in the simulations, and the hydrofoil surface is a no-slip wall. The domain is kept identical for all two-dimensional cases so that changes in the calculated cavitation behavior are attributable to the control configuration rather than to changes in the external boundaries.

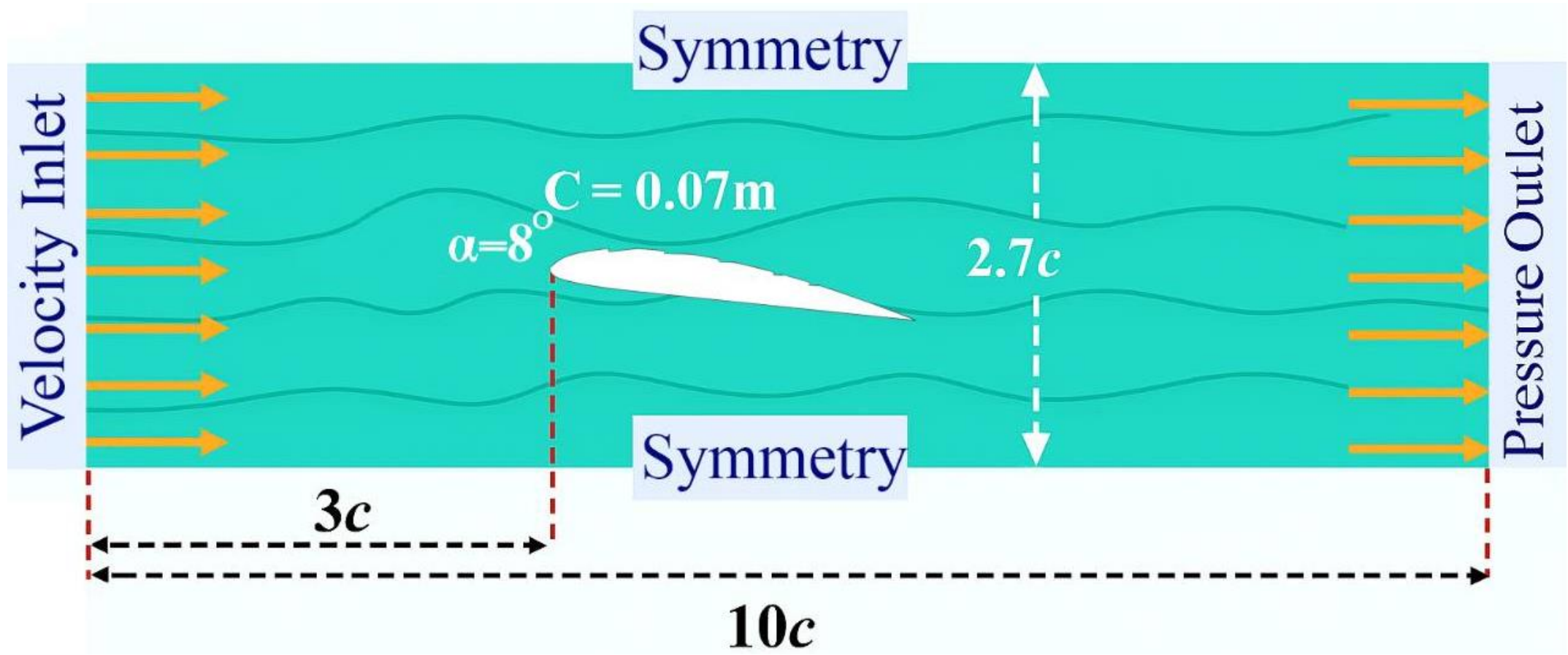

Fig. 2. Computational domain and boundary-condition layout used for the two-dimensional simulations.

### 3.2. Boundary and operating conditions

The baseline operating condition follows the established Clark-Y cloud-cavitation benchmark of Roohi et al. [6]. The free-stream velocity is $U_\infty$ = 10 m/s, the chord Reynolds number is Re = $7\times10^5$, and the prescribed cavitation number is σ = 0.8. The hydrofoil is treated as a stationary no-slip wall, whereas the upper and lower domain boundaries are specified as symmetry boundaries. The simulations are initialized using Standard Initialization from the inlet, and the initial vapor volume fraction is set to zero.

For clarity, the cavitation number is defined as $\sigma = \frac{p_\infty - p_v}{0.5\,\rho_l\,U_\infty^2}$ where $p_\infty$ denotes the reference pressure used in the benchmark cavitation-number definition, $p_v$ is the saturation vapor pressure, and $\rho_l$ is the liquid density. These reference pressures are used only in the definition of σ and are not inferred from the numerical value entered in the Fluent pressure-outlet gauge field. Surface pressure is reported through $C_p = \frac{p - p_\infty}{0.5\,\rho_l\,U_\infty^2}$ The hydrodynamic coefficients are defined using the chord and unit span in the conventional manner: $C_L = \frac{L}{0.5\,\rho_l\,U_\infty^2\,c}$ and $C_D = \frac{D}{0.5\,\rho_l\,U_\infty^2\,c}$

These definitions are used consistently for the baseline and controlled cases.

### 3.3. Tangential micro-jet configuration

The slot width is fixed at w = 0.57 mm for all controlled cases. A uniform liquid-injection velocity of $U_{jet}$ = 2.5 m/s is prescribed at the slot, corresponding to $U_{jet}/U_\infty = 0.25$. For the two-dimensional unit-span formulation, the same information can be expressed through the jet momentum coefficient $C_\mu = \rho_l\, U_{jet}^2\, w/[0.5\, \rho_l\, U_\infty^2\, c] = 2(w/c)(U_{jet}/U_\infty)^2 \approx 1.02 \times 10^{-3}$, where the injected liquid and free stream have the same density and the unit span cancels between jet and reference areas. The injection direction is aligned tangentially with the local suction-surface direction. Because the jet velocity is explicitly imposed as a boundary condition, the present model represents prescribed tangential momentum injection rather than a self-driven pressure-fed channel. Consequently, the comparisons isolate the influence of chordwise injection location at fixed $U_{jet}/U_\infty$ and $C_\mu$; they do not evaluate energetic cost, internal-channel losses, or self-regulation of a passive fluidic device.

The five controlled locations cover the forward cavity-development region, intermediate chord positions, and the downstream recovery region (Fig. 1). Holding w, $U_{jet}/U_\infty$, $C_\mu$, and injection direction fixed permits the response to be interpreted primarily in terms of where the imposed momentum enters relative to the evolving attached cavity and the near-wall momentum deficit. Previous investigations have shown that jet position, angle, and momentum input can strongly influence cavity response [10-15], which motivates isolating chordwise position in the present comparison.

### 3.4. Computational mesh and spatial resolution

A structured multi-block mesh is used for the two-dimensional calculations, with local refinement concentrated where strong velocity, pressure, and phase-fraction gradients are expected. Fig 3 separates the principal refinement regions: the leading edge, where the pressure decreases rapidly and the attached cavity originates; the trailing edge and wake; the injection-slot neighborhood; and the global domain. This targeted distribution is intended to preserve near-wall and interface resolution without unnecessarily refining the entire far field.

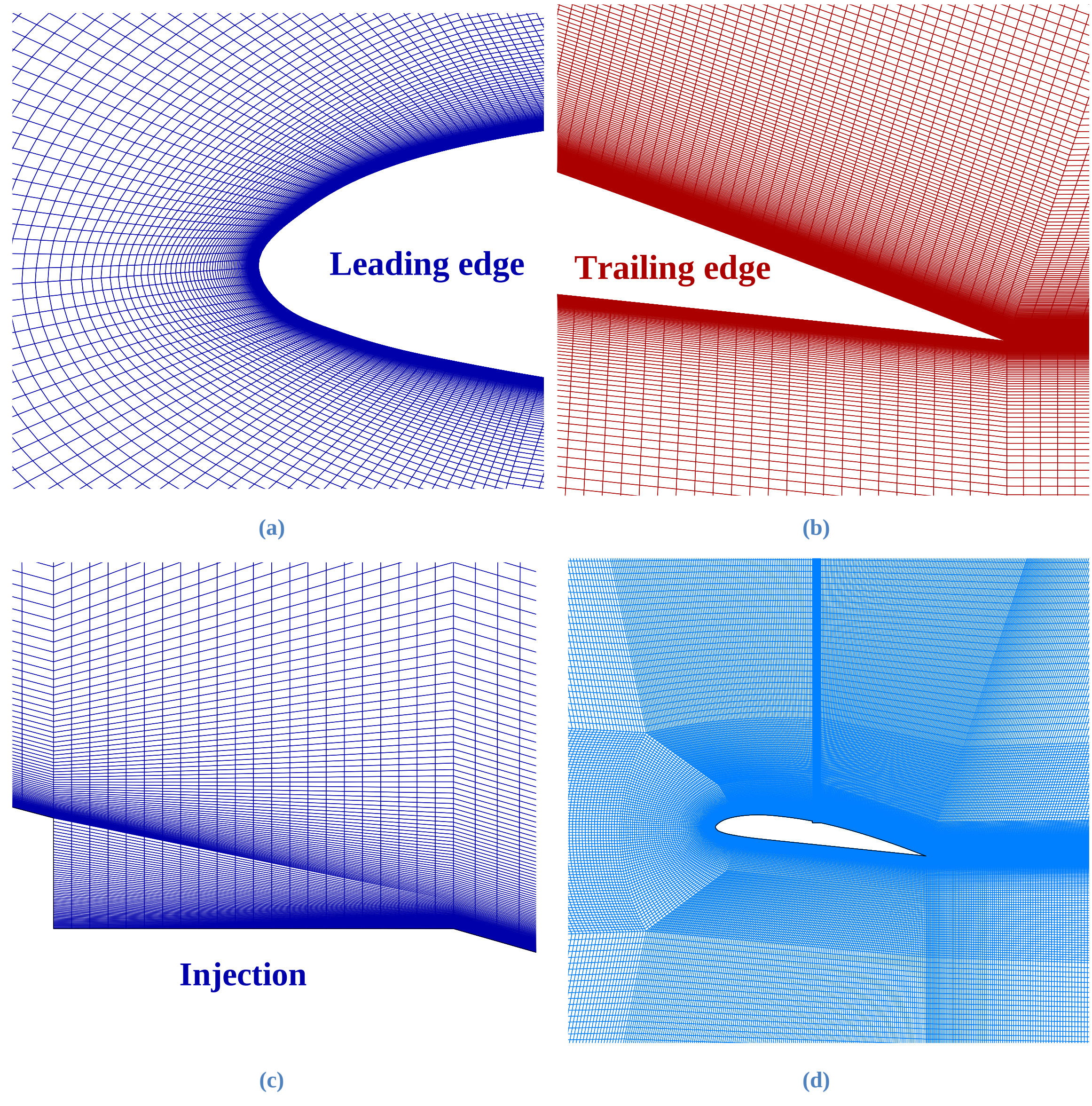

Fig. 3. Structured multi-block mesh topology around the two-dimensional Clark-Y hydrofoil: (a) leading-edge refinement, (b) trailing-edge refinement, (c) local refinement around the injection slot, and (d) global mesh distribution.

Spatial sensitivity is assessed using approximately $1.5 \times 10^5$, $2.4 \times 10^5$, and $3.1 \times 10^5$ cells. The comparison metric is the surface pressure-coefficient distribution shown in Fig. 4. The medium and fine meshes produce closely similar $C_p$ profiles over most of the chord, whereas the coarsest mesh shows larger deviations. The medium mesh, containing approximately 2.4 × 105 cells, is therefore used for the parametric calculations as a compromise between spatial resolution and computational cost. Importantly, Fig. 4 is a $C_p$-based grid-sensitivity assessment rather than evidence that every instantaneous or integral quantity is fully grid-independent. More rigorous solution-verification procedures would require systematic uncertainty estimation for each quantity of interest [33,34].

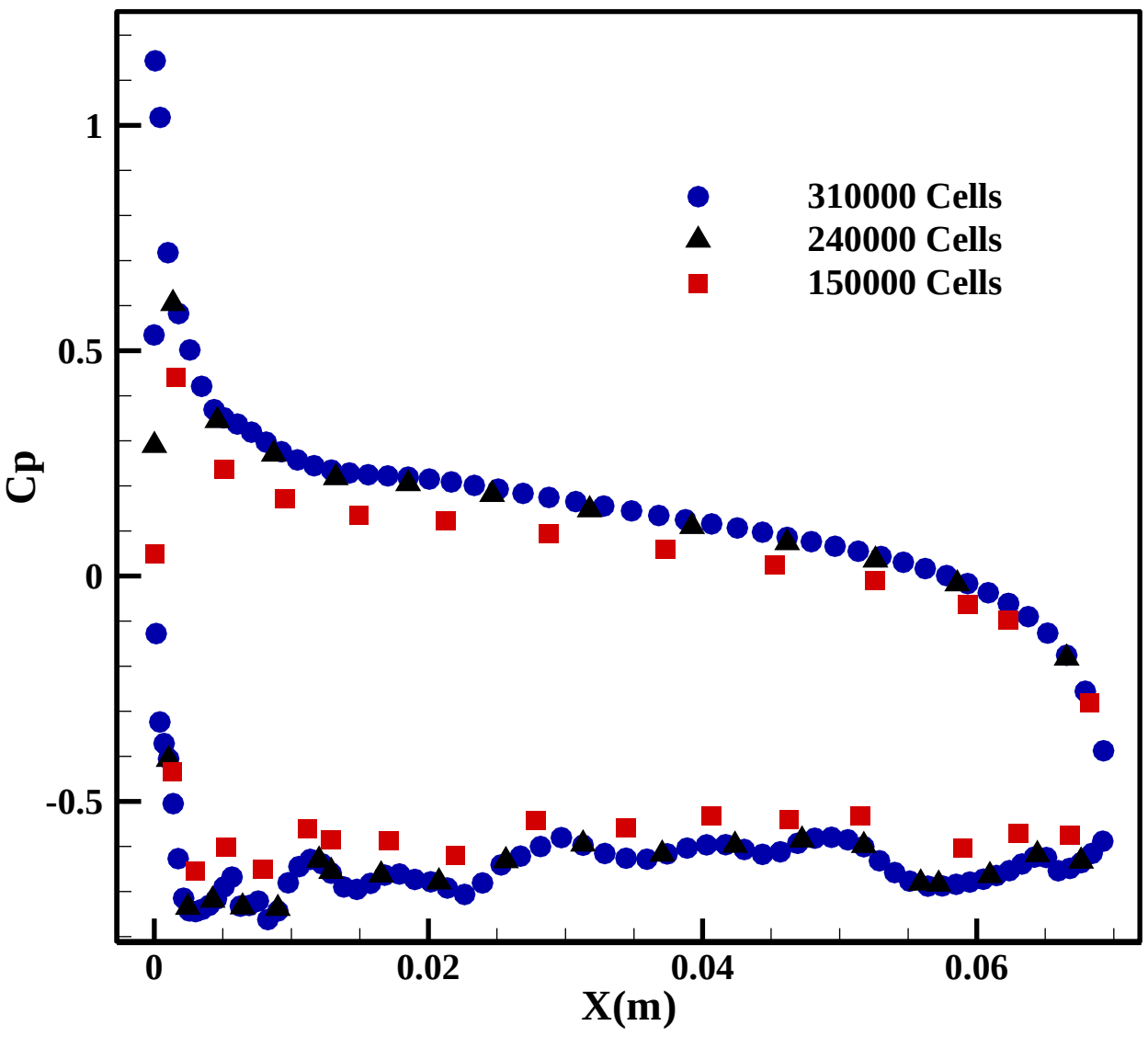

Fig. 4. Grid sensitivity assessment based on the surface pressure-coefficient distribution for three mesh resolutions.

Near-wall resolution is evaluated through the wall distance $y^+$. As shown in Fig. 5, $y^+$ remains close to unity over most of the suction surface, with a localized increase near the injection region. The distribution supports direct resolution of the near-wall region over most of the chord while also identifying the slot neighborhood as the most demanding location. This local departure should be kept in mind when interpreting fine-scale turbulence close to the injection interface, particularly because numerical predictions remain sensitive to spatial resolution and turbulence closure [35,36].

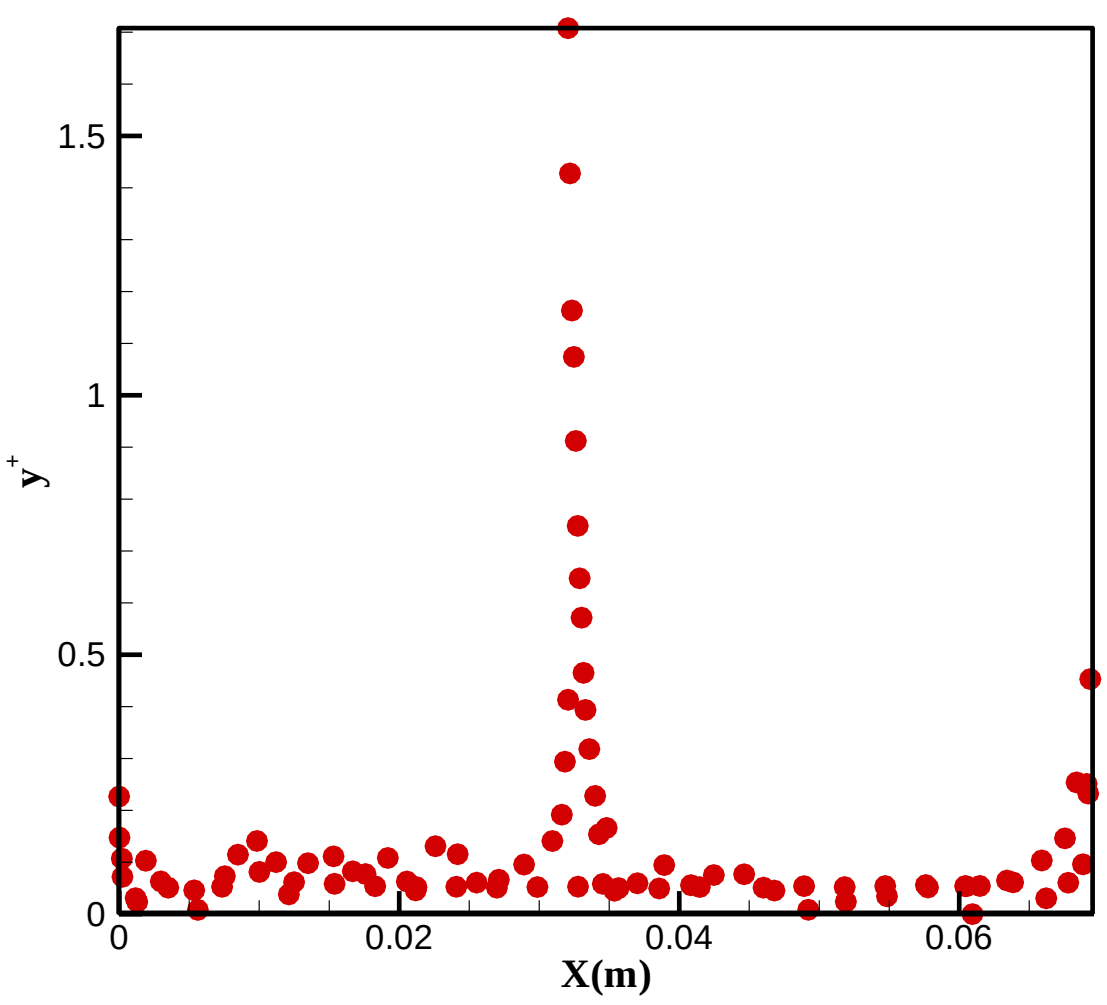

Fig. 5. Time-averaged wall-distance distribution $y^+$ along the hydrofoil suction surface for Case 3.

### 3.5. Temporal discretization and numerical schemes

The transient calculations are performed in ANSYS Fluent using a pressure-based formulation. Pressure-velocity coupling is handled with the Pressure-Implicit with Splitting of Operators (PISO) algorithm. The spatial discretization uses second-order-accurate formulations for the transported variables, including bounded central differencing where appropriate for the LES momentum equations. An adaptive time-stepping strategy is employed rather than a single fixed physical time step; the physical time-step size is automatically adjusted during the transient calculation so that the global Courant-Friedrichs-Lewy number remains below 0.45.

At each adaptive time step, convergence is monitored using normalized residuals, with a target below $1 \times 10^{-7}$ for the solved equations. The same numerical settings are retained for all baseline and controlled cases. To exclude the initial transient development from the force statistics, the first complete cavitation cycle is omitted. The reported lift and drag coefficients are cycle-averaged quantities obtained by averaging the instantaneous force histories over the complete second and third cavitation cycles. Because cavitation-model and turbulence-model choices can influence quantitative predictions [20,21,35,37,38], the interpretation focuses on consistent case-to-case trends within the adopted framework rather than on claiming model-independent values. The surface-pressure distributions reported in Figs. 6 and 13 were averaged over the same complete second and third cavitation cycles, with the first cycle excluded from the averaging.

Following common cavitating-flow post-processing practice [42], the stored ANSYS Fluent field is reported here using the conventional turbulent-kinetic-energy notation k. Because a separate resolved/subgrid energy decomposition is not required for the present comparison, Fig. 10 is interpreted in terms of relative spatial distribution and normalized intensity rather than as a resolved-to-subgrid energy budget. For post-processing of the turbulent kinetic energy fields, a single common normalization factor is used for every case and sampled instant: $k^* = k/k_{max,global}$, where $k_{max,global}$ is the maximum turbulent kinetic energy evaluated over all investigated configurations and all sampled times included in the TKE comparison. This global normalization permits direct comparison of the relative TKE intensity among the different cases and times.

The available lift- and drag-coefficient histories were additionally post-processed to characterize unsteady loading independently of the cycle-averaged force statistics reported in Table 2. Because the CFD calculations used adaptive physical time stepping, the cleaned force signals were interpolated onto a uniform post-processing grid with $\Delta ts = 2 \times 10^{-6}$ s; this interval is a resampling interval and not the CFD solver time step. The record mean was removed before RMS evaluation. For spectral estimation, each resampled signal was linearly detrended, multiplied by a Hann window, and analyzed using a full-record periodogram. Because the available records have finite and case-dependent durations, the RMS values are treated as record-based fluctuation metrics and the spectra are used only for relative force-response comparison. Precise cavity-shedding frequencies are not inferred from these force spectra in the absence of an independent cavity-based time signal.

### 3.6. Baseline validation

Validation is restricted to the uncontrolled baseline at the same principal operating condition used in the parametric study. The baseline geometry and operating regime correspond to the Roohi-Zahiri Clark-Y benchmark [6] ($c$ = 70 mm, $\alpha$ = 8°, $U_\infty$ = 10 m/s, $Re$ = 7 × $10^5$, and $\sigma$ = 0.8), while the present Fluent implementation retains its own solver and turbulence-model formulation. Fig 6 compares the cycle-averaged surface pressure coefficient with the numerical benchmark of Roohi et al. [6]. The present calculation reproduces the main pressure distribution, including the strong leading-edge suction, the broad cavitating pressure plateau, and the downstream pressure recovery. The validation results show good agreement with the corresponding results reported in [6].

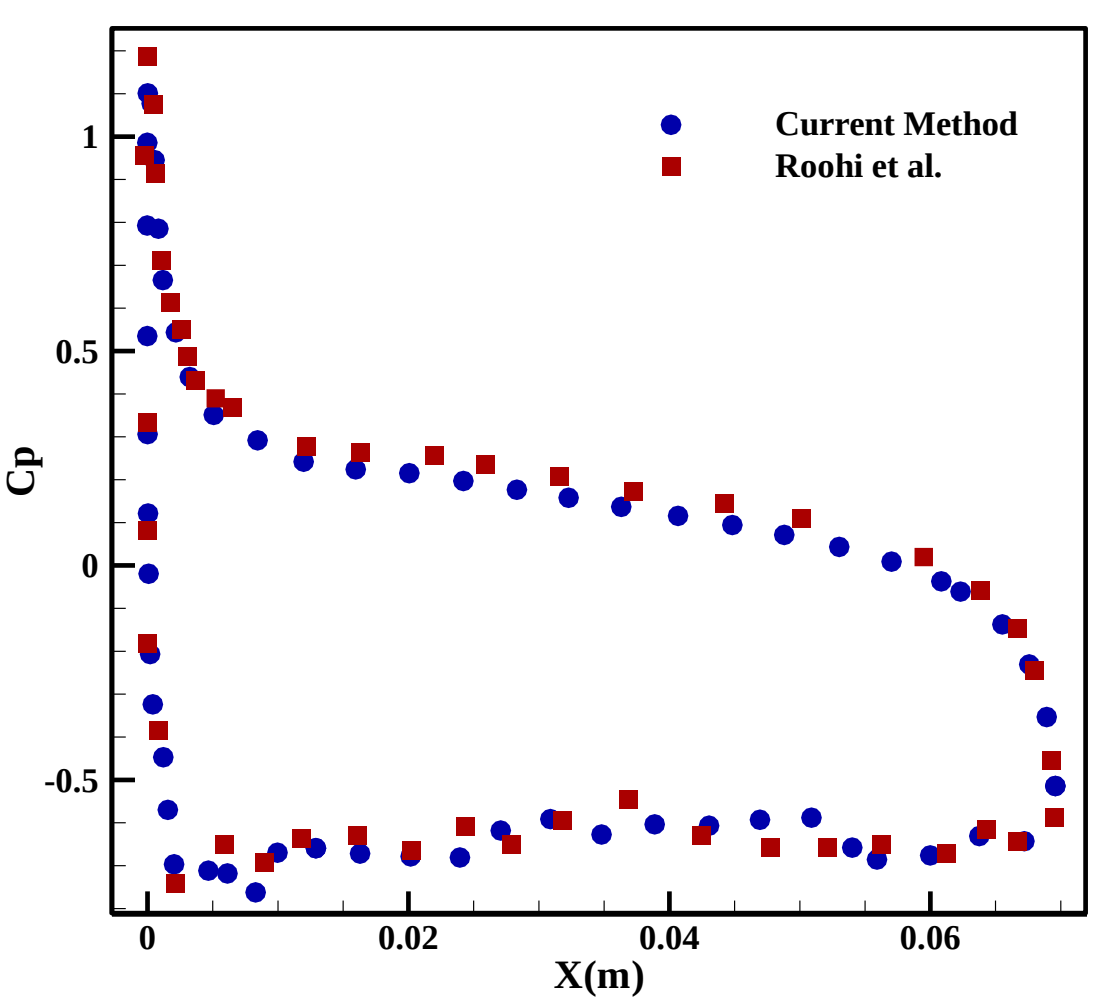

Fig. 6. Comparison of the cycle-averaged surface pressure coefficient with the numerical benchmark of Roohi et al. [6].

A second comparison is provided through the cycle-averaged lift and drag coefficients in Table 1. The present $C_L$ = 0.78 differs by approximately 2.6% from the experimental value of 0.76 reported for the same benchmark condition in Roohi et al. [6], whereas $C_D$ = 0.137 is approximately 14.1% above the corresponding experimental value of 0.120. The lift prediction is therefore close to the benchmark, but the larger drag discrepancy is retained explicitly as a quantitative limitation when interpreting differences among the controlled cases. The current drag value is close to the numerical result reported by Roohi et al. [6], and the overall benchmark condition is consistent with the experimental/numerical Clark-Y literature [7,32].

Table 1. Comparison of cycle-averaged lift and drag coefficients with the Roohi-Zahiri Clark-Y benchmark data.

| Source | $C_L$ | $C_D$ |
|---|---|---|
| Present numerical study | 0.78 | 0.137 |
| Roohi et al. [6] | 0.78 | 0.140 |
| Experiment reported in Roohi et al. [6] | 0.76 | 0.120 |
| Deviation from experiment (%) | 2.6 | 14.1 |

### 3.7. Representative three-dimensional simulation

The primary location sweep is two-dimensional. To illustrate flow features that cannot exist in a strictly planar representation, one additional three-dimensional calculation is performed for Case 3, with the slot located at x/c = 0.30. The two-dimensional geometry is extruded over a spanwise distance of 21 mm, and symmetry conditions are applied at the two spanwise side boundaries. The inlet velocity, outlet gauge-pressure specification, cavitation number, turbulence model, cavitation model, and jet condition are maintained consistently with the corresponding two-dimensional case.

Fig 7 shows the three-dimensional mesh. Local refinement is concentrated near the hydrofoil, injection region, and cavitation-prone suction-side flow. The final mesh contains approximately $1.57 \times 10^6$ cells. Previous three-dimensional LES and Lagrangian investigations have demonstrated the importance of spanwise and side-entrant structures in unsteady cavitating hydrofoil flows [39,40], providing the physical motivation for this representative comparison.

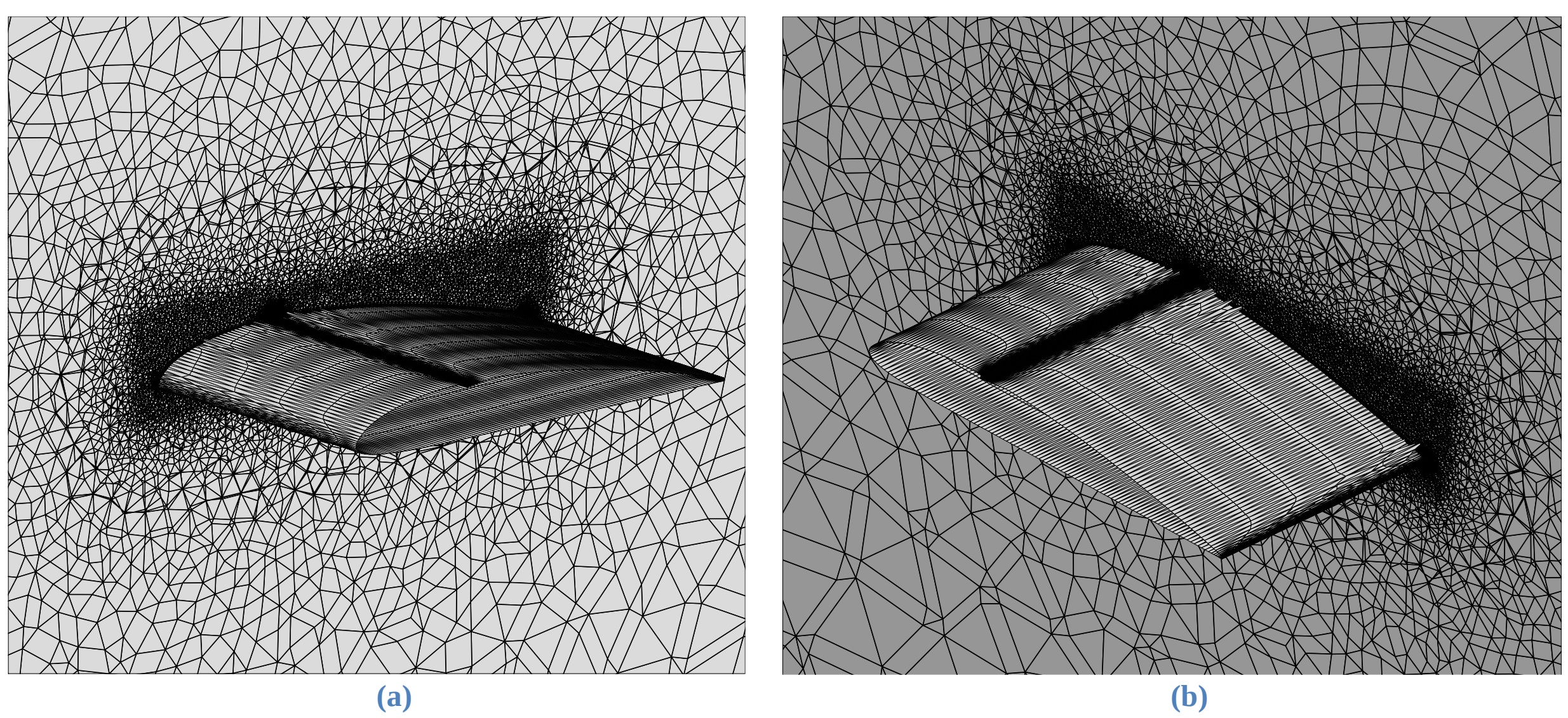

Fig. 7. Three-dimensional computational mesh and local refinement around the extruded Clark-Y hydrofoil.

## 4. Results and discussion

The results are organized from qualitative dimensionality effects to the systematic location sweep and then to integrated hydrodynamic measures. Direct observations from the simulations are distinguished from mechanistic interpretation, and the latter is related to established cavitation physics where supporting evidence is available. Because the two-dimensional cases are the basis of the location sweep, the three-dimensional result is used to contextualize, rather than replace, the parametric analysis.

### 4.1. Representative comparison of two- and three-dimensional cavity structures

Fig 8 presents representative vapor structures from the planar two-dimensional solution and the complementary three-dimensional Case 3 calculation. The figure is used to address a specific question: what flow features are structurally excluded by the two-dimensional approximation? In the planar calculation, the vapor structures are necessarily coherent across the absent spanwise direction. The three-dimensional visualization, in contrast, exhibits spanwise deformation, nonuniform cavity closure, and fragmentation into structures that are not constrained to a single plane. These differences are consistent with previous three-dimensional studies in which side-entrant motion, vortex stretching, and spanwise cavity deformation contributed to cloud breakup [4,25,39,40].

The comparison should not be interpreted as a quantitative validation of the two-dimensional location ranking. Rather, it shows why the 2D sweep is best regarded as a controlled parametric framework: it preserves the principal chordwise coupling between pressure, cavity growth, and imposed momentum while omitting genuine three-dimensional turbulent pathways. The 3D result therefore strengthens the physical context of the study while also defining an important limitation of the broader location comparison.

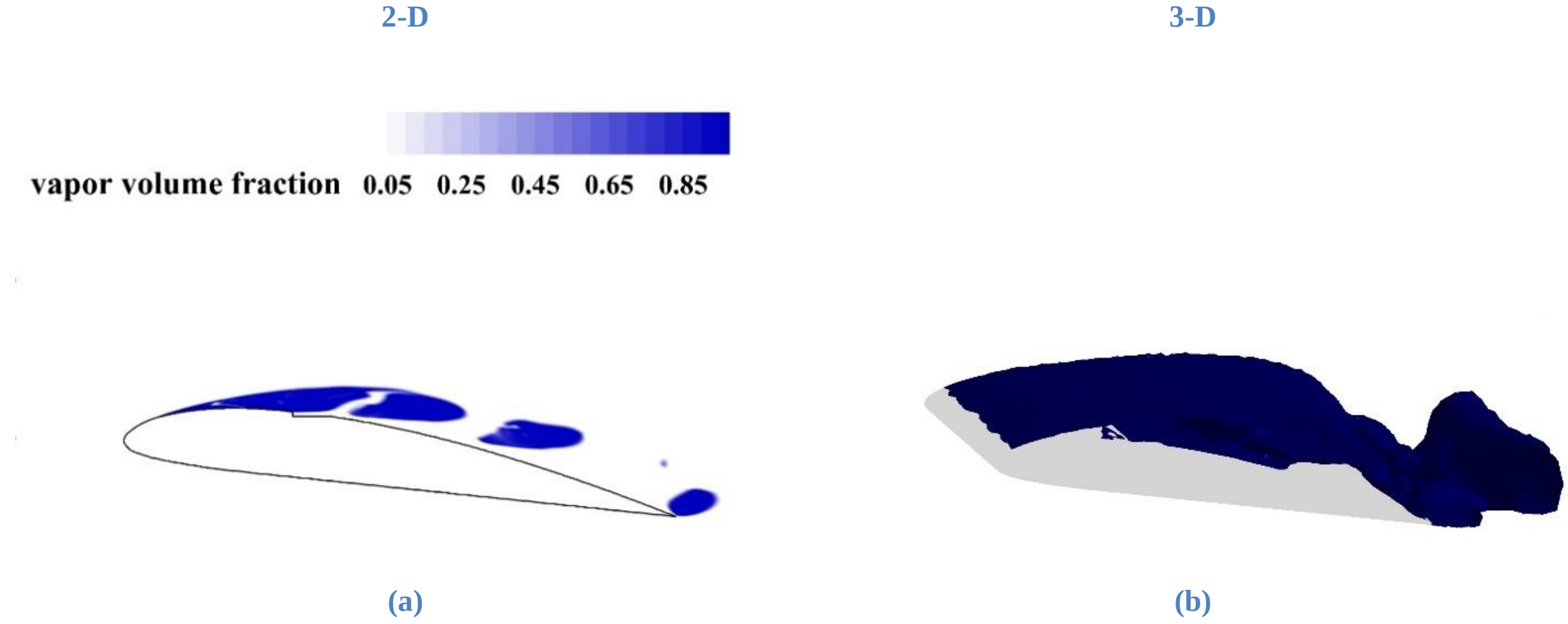

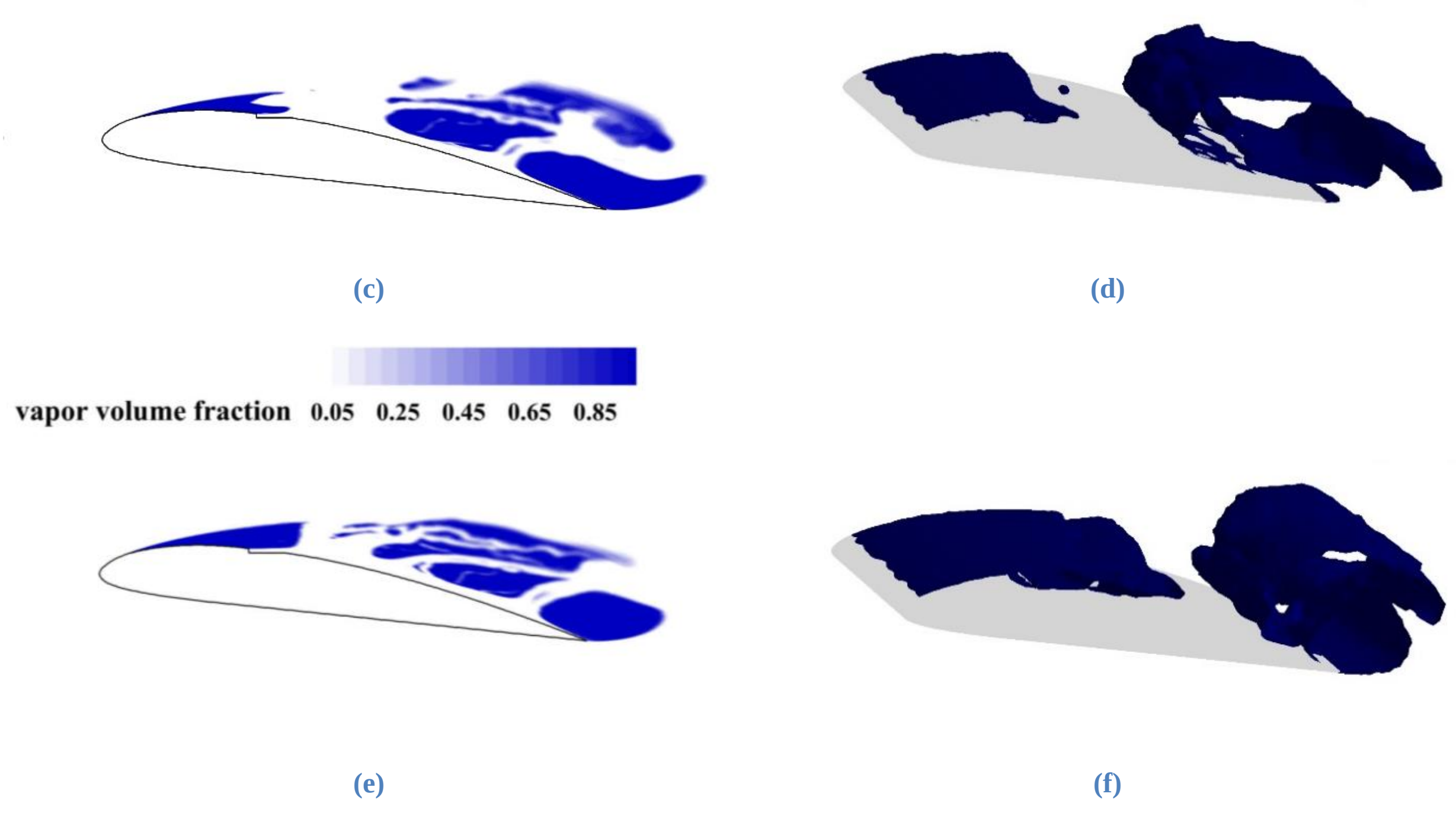

Fig. 8. Representative comparison of vapor structures obtained from the two-dimensional calculation (a, c, e) and the corresponding three-dimensional Case 3 calculation (b, d, f) under the same operating condition.

### 4.2. Effect of chordwise jet location on transient vapor evolution

The location dependence of the cavitating field is first examined through the vapor-volume-fraction sequence in Fig. 9. The snapshots are reported at common physical times from 20 to 50 ms. Because the available post-processing does not establish phase synchronization among cases with potentially different shedding periods, equal physical times are treated here as temporal samples rather than as equivalent phases of a universal shedding cycle. This distinction is important when comparing instantaneous cavity size or the apparent stage of detachment.

Despite this limitation, a consistent spatial pattern emerges across the sequence. The baseline develops an attached vapor region over the forward suction surface and periodically releases large detached structures into the wake. Introducing the jet at $x/c = 0.15$ changes the cavity close to the region where the attached sheet is still developing, whereas locations farther downstream interact with a cavity that has already occupied a larger fraction of the chord. The forward cases therefore modify the cavity earlier in its streamwise evolution; mid-chord and rearward cases primarily alter the cavity after substantial upstream development has already occurred. This is consistent with previous active-injection studies showing that the response depends strongly on where the jet intersects the cavity and the near-wall liquid layer [10,12–15].

The sequence also shows that no controlled case can be characterized by a single instantaneous image. Large vapor structures appear at different physical times in different cases, emphasizing that a location can alter both cavity morphology and temporal organization. Consequently, the evidence in Fig. 9 is interpreted together with the cycle-averaged pressure and force data rather than used by itself to claim complete suppression. Experimental measurements of partial/cloud cavitation likewise show that vapor fraction and cavity topology vary strongly across the closure and shedding process [41].

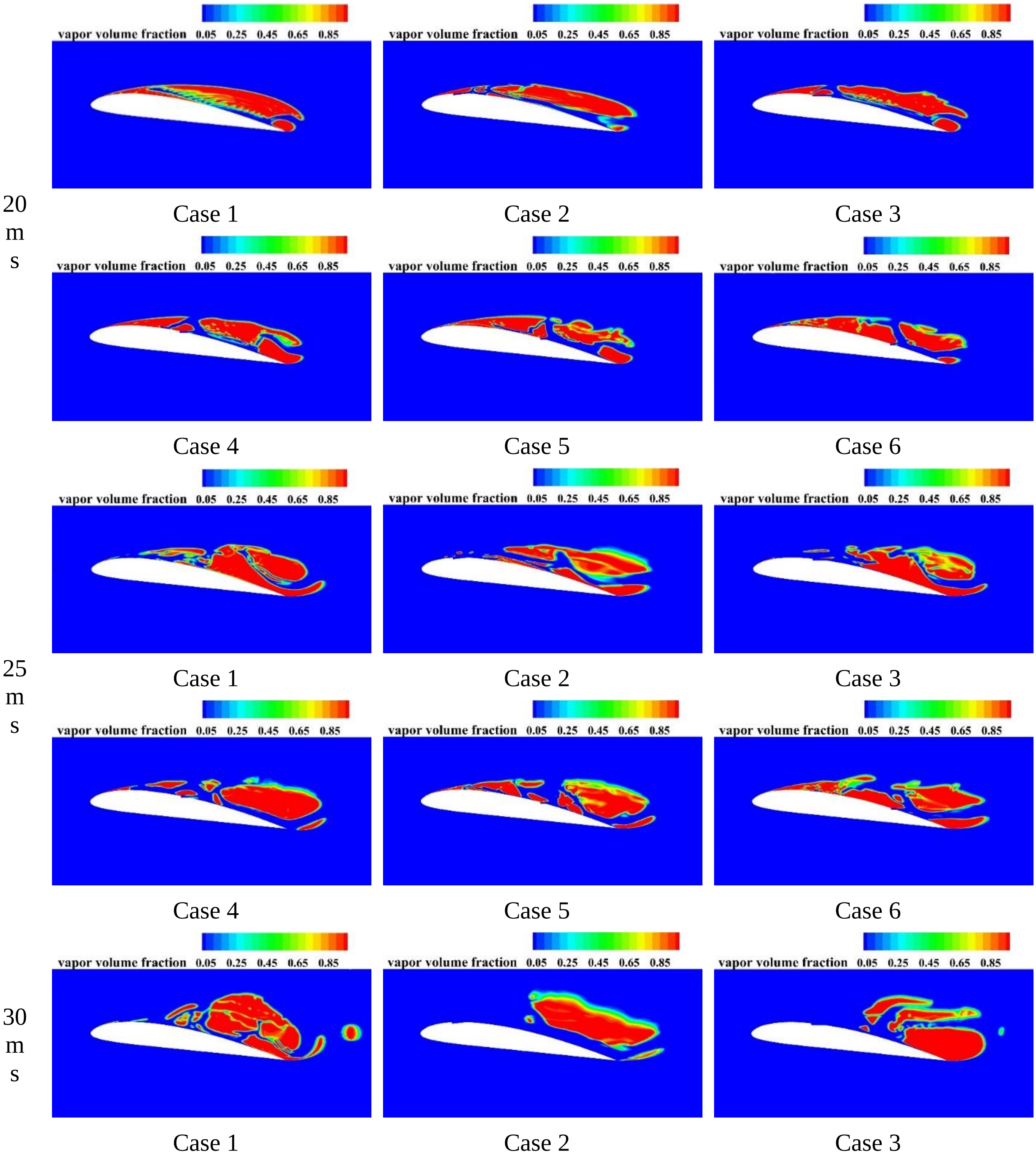

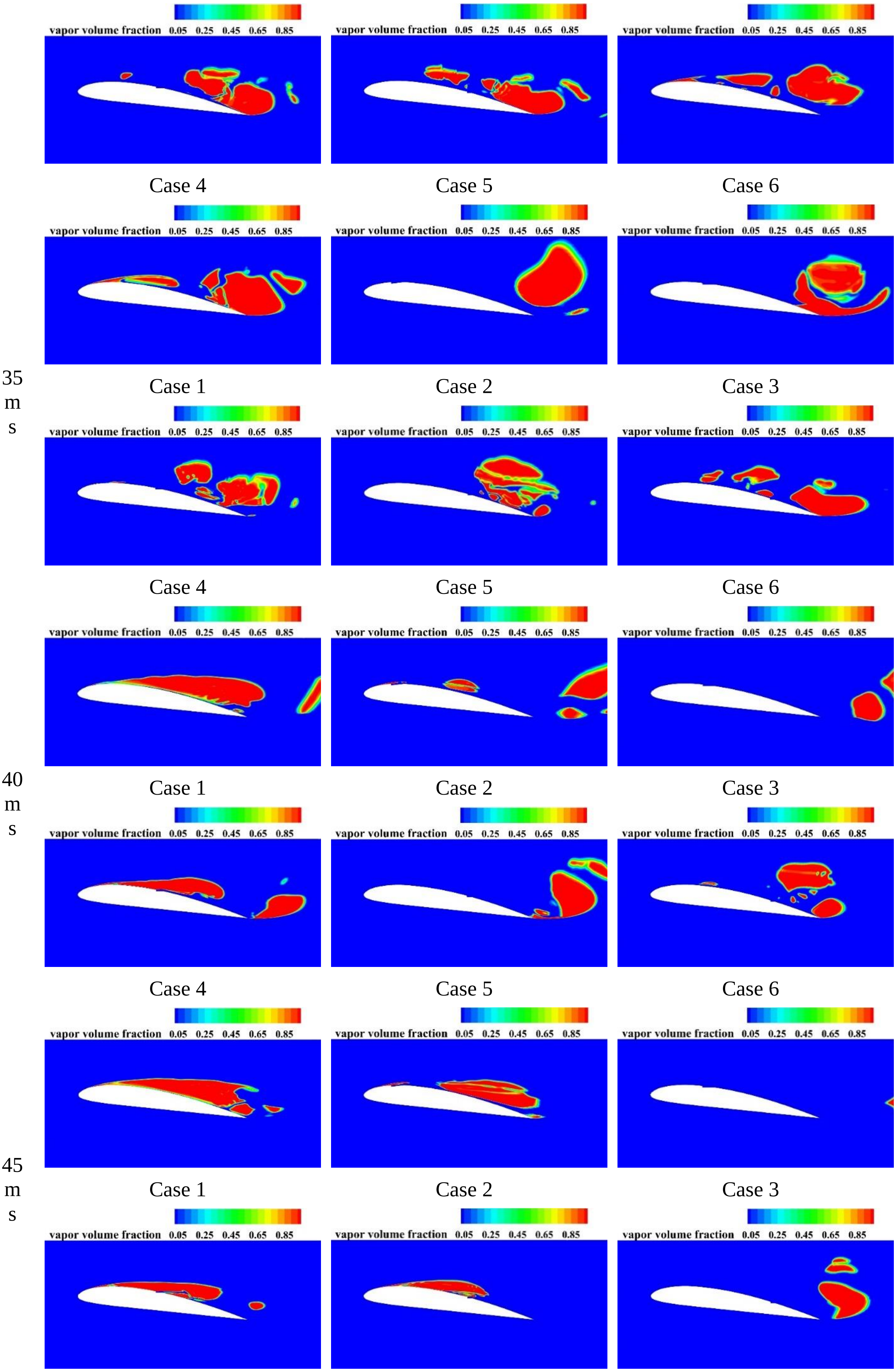
vapor volume fraction 0.05 0.25 0.45 0.65 0.85
Case 4
Case 5
Case 6
Case 1
Case 2
Case 3
35 ms
Case 4
Case 5
Case 6
Case 1
Case 2
Case 3
40 ms
Case 4
Case 5
Case 6
Case 1
Case 2
Case 3
45 ms

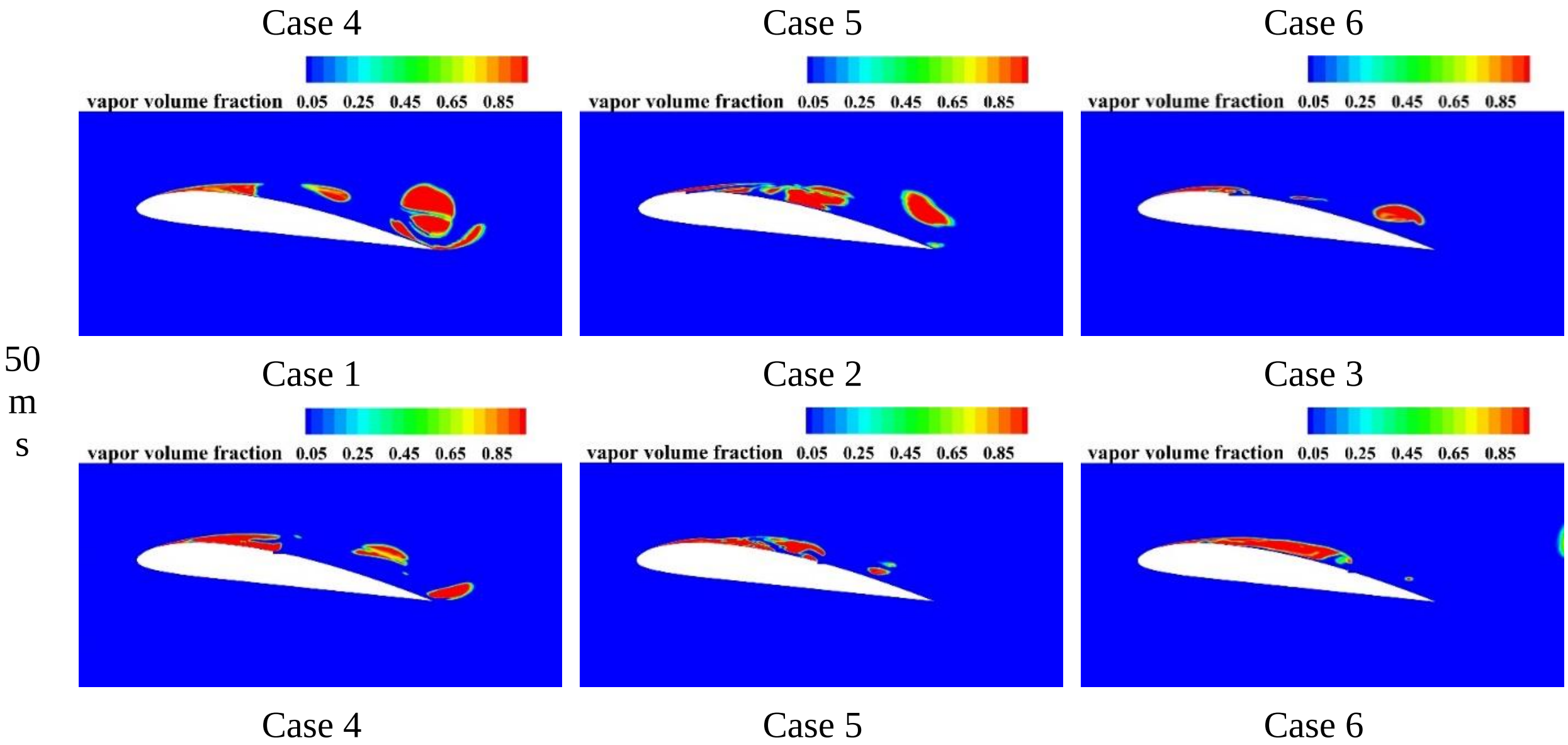

Fig. 9. Temporal evolution of vapor volume fraction for the uncontrolled baseline and five prescribed tangential micro-jet locations from t = 20 to 50 ms.

### 4.3. Turbulent and velocity-field response

Fig 10 examines the normalized turbulent kinetic energy (TKE) field at 30, 40, and 50 ms. A single global maximum is used for every panel, such that $k^* = k/k_{max,global}$, with $k_{max,global}$ obtained from all investigated configurations and sampled instants. Because the same normalization factor is applied throughout, relative TKE intensity can be compared directly among cases and times. In the baseline, elevated TKE is concentrated around the cavity/shear-layer region and extends into the downstream wake as the vapor structure evolves. This spatial association is expected in sheet/cloud cavitation because cavity deformation, shear-layer roll-up, and vortical motion are strongly coupled [5,29,42]. The controlled cases redistribute these high-TKE regions rather than simply removing them. Forward injection tends to shift the strongest activity away from a long continuous near-wall region, while mid- and rear-chord injection introduces localized turbulent activity around and downstream of the slot.

The TKE contours are therefore most useful as indicators of where turbulent fluctuations are concentrated. They should not be interpreted directly as erosion, acoustic intensity, or collapse-shock strength. Erosion-risk inference requires dedicated impact/erosion modeling [43], whereas collapse-pressure or shock-wave quantification requires a compressibility-aware treatment [3]. Within this limitation, Fig. 10 supports the interpretation that chordwise injection location changes the coupling between the near-wall flow, cavity interface, and downstream turbulent wake.

A more detailed comparison of Fig. 10 reveals a clear temporal and location-dependent redistribution of the turbulent activity. At t = 30 ms, the baseline case exhibits elevated normalized

TKE primarily around the suction-side cavity/shear-layer region and near the downstream cavity closure, whereas the controlled cases modify both the extent and position of these high-k zones. The forward injection case at x/c = 0.15 produces a broader redistribution of turbulent activity over the forward-to-middle portion of the suction side, consistent with the jet interacting with the cavity while it is still developing. As the injection location is shifted downstream, particularly for Cases 4-6, the changes become more localized around the cavity-closure and downstream regions. At t = 40 ms, the high-k structures have evolved further into the wake, and the different configurations exhibit distinct downstream patterns, indicating that the jet location affects not only the local turbulence level but also the subsequent transport of turbulent structures. By t = 50 ms, the spatial distribution again changes considerably among the cases: some configurations exhibit relatively weak and elongated downstream activity, whereas Case 6 shows a pronounced localized high-k region downstream of the hydrofoil. These temporal variations demonstrate that tangential injection does not simply suppress turbulent activity uniformly; rather, changing the chordwise jet location modifies where and when turbulence is generated and transported during the cavity-evolution process.

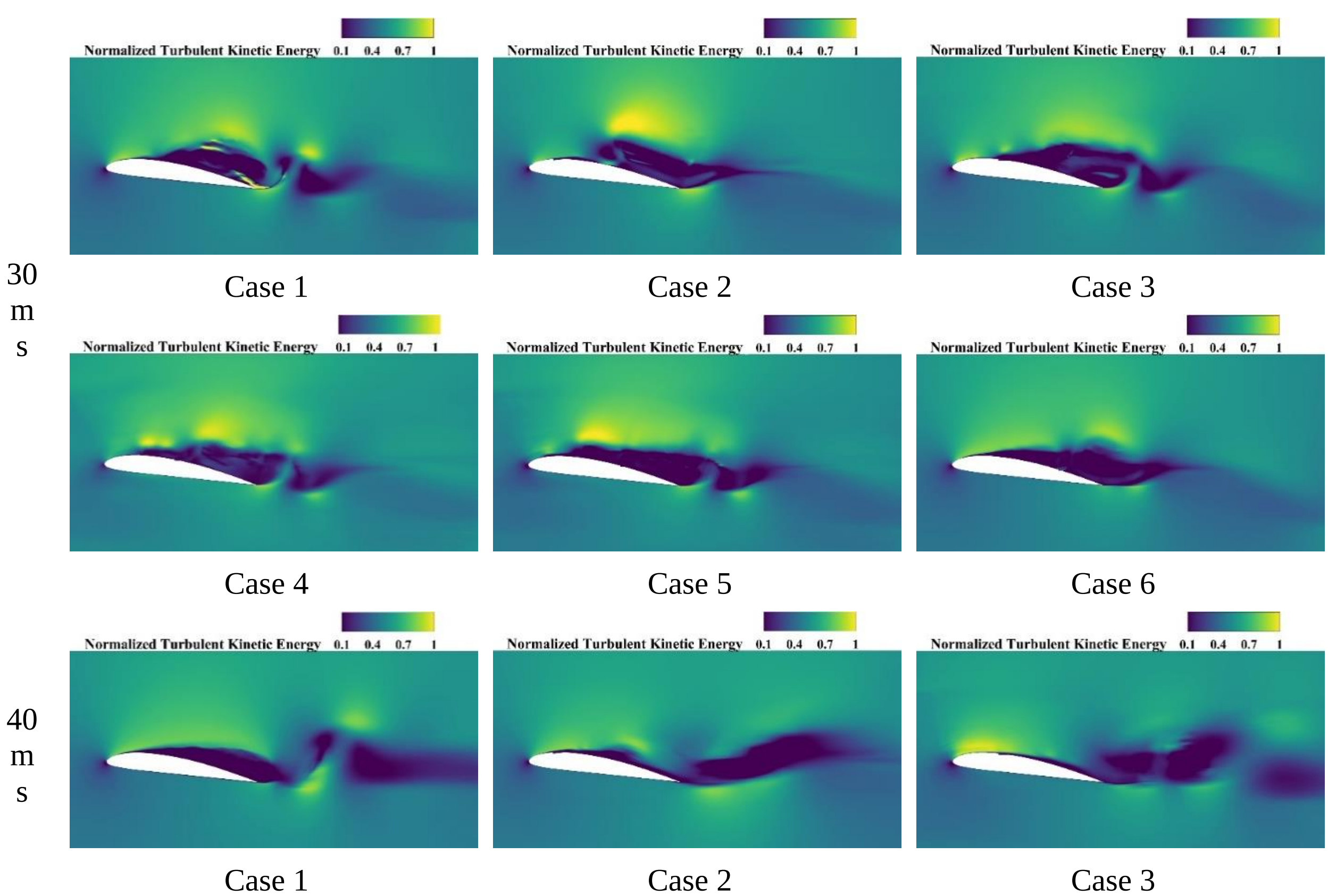

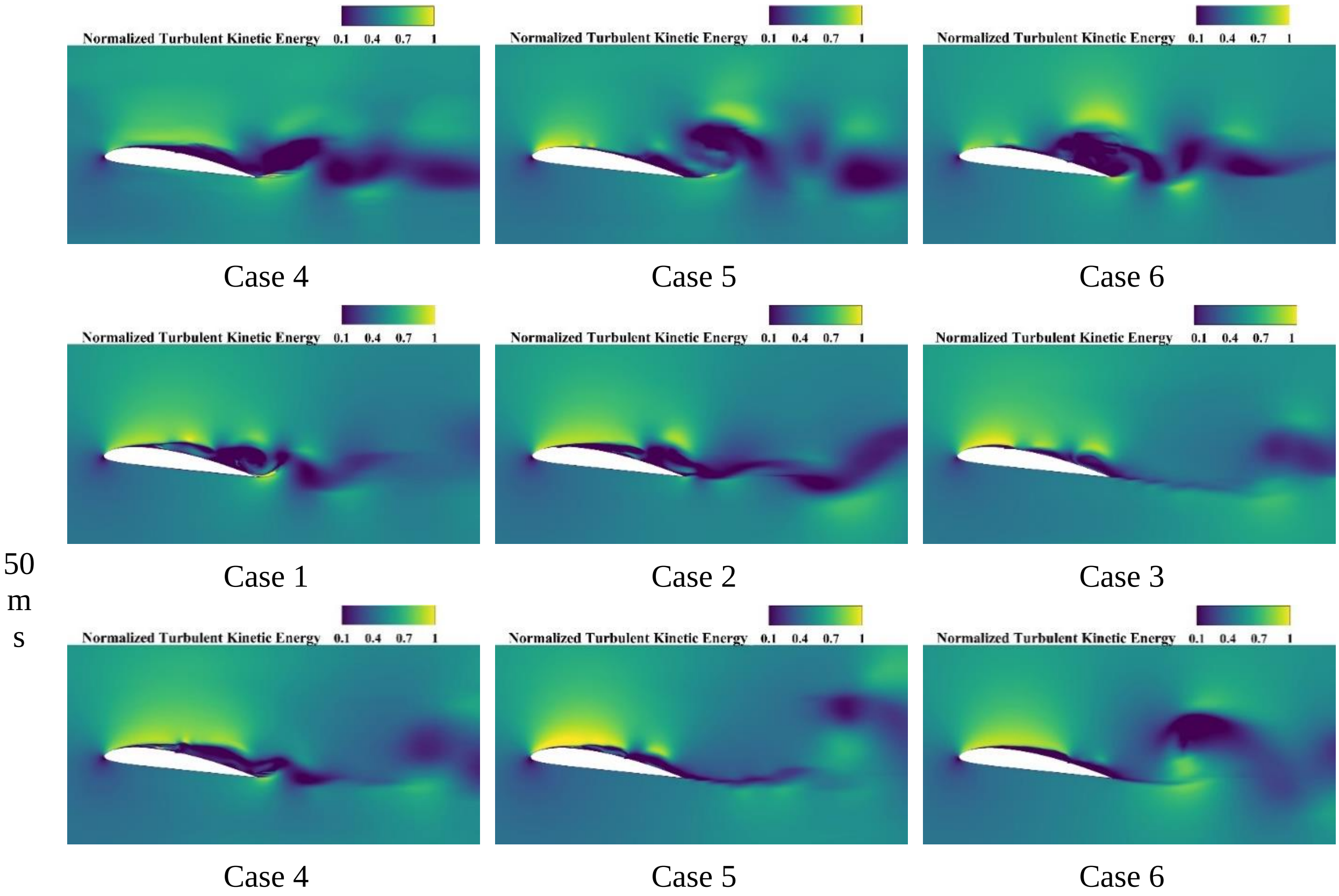

Fig. 10. Temporal evolution of normalized turbulent kinetic energy for the baseline and controlled configurations at t = 30, 40, and 50 ms. All panels use the same global normalization, $k^* = k/k_{max,global}$, where $k_{max,global}$ is the maximum TKE over all investigated cases and sampled instants.

The corresponding velocity-magnitude fields in Fig. 11 provide a complementary view of the momentum distribution. The baseline contains a pronounced low-velocity region associated with the cavitating near-wall flow and a broad downstream velocity deficit at several sampled instants. When injection is placed near the forward chord, the additional tangential momentum modifies the near-wall velocity field before the cavity reaches the downstream half of the foil. The influence becomes progressively more localized as the slot is moved toward the trailing edge, because a substantial upstream momentum deficit has already formed before the injected fluid is introduced.

The velocity-magnitude plots do not provide flow direction and therefore cannot, by themselves, prove elimination or blockage of an upstream-moving re-entrant jet. They do, however, show systematic changes in the magnitude and spatial extent of the near-wall and wake deficits. The combined behavior of Figs. 9-11 is thus consistent with a location-dependent modification of the liquid momentum available beneath and around the cavity, rather than with a universal suppression mechanism.

A more detailed examination of Fig. 11 shows that the velocity-magnitude field evolves differently with both time and jet location. At t = 30 ms, the baseline case exhibits a pronounced low-velocity region over the suction side and immediately downstream of the cavity-closure region, together with a relatively broad wake deficit. The controlled cases modify this structure to different degrees. For the forward injection case at x/c = 0.15, the added tangential momentum affects the velocity field over a relatively large portion of the suction side and alters the downstream deficit from an early stage of cavity development. As the slot is moved toward the mid- and rear-chord positions, the upstream part of the low-velocity region becomes increasingly similar to the baseline, while the strongest modifications become concentrated around and downstream of the injection location. At t = 40 ms, the differences among the cases become more pronounced. The baseline still contains an extended near-wall velocity deficit and a non-uniform wake, whereas Cases 2 and 3 show substantial redistribution of the low-velocity zones over the suction-side and downstream regions. For Cases 4-6, the upstream velocity field is less affected, but distinct localized low-velocity structures develop farther downstream, indicating that rearward injection mainly modifies the later stages of the cavity-wake evolution. By t = 50 ms, the velocity deficits have been convected farther downstream and the spatial organization differs markedly among the configurations. Some cases, particularly the forward and intermediate injection configurations, exhibit a comparatively elongated but weaker downstream deficit, while the rearward cases show more localized separated low-velocity regions in the wake; Case 6, in particular, displays a pronounced downstream low-velocity structure. Overall, the temporal sequence demonstrates that the prescribed jet does not produce a uniform increase in local velocity throughout the flow field. Instead, its chordwise position determines the streamwise location and stage of cavity evolution at which the near-wall momentum deficit and downstream wake are modified. These observations are consistent with a location-dependent redistribution of momentum, although the velocity-magnitude contours alone do not provide sufficient information to determine the direction of the local flow or to demonstrate complete suppression of the re-entrant jet.

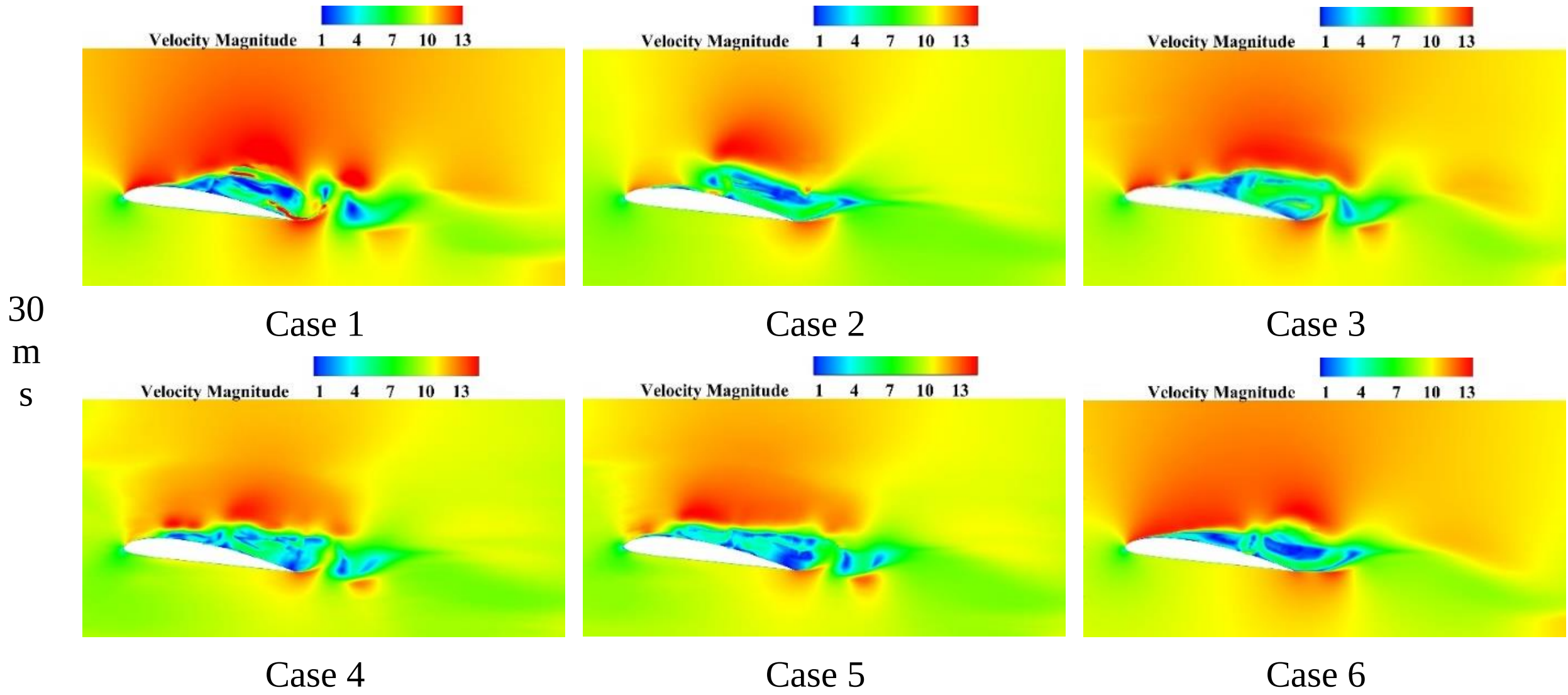

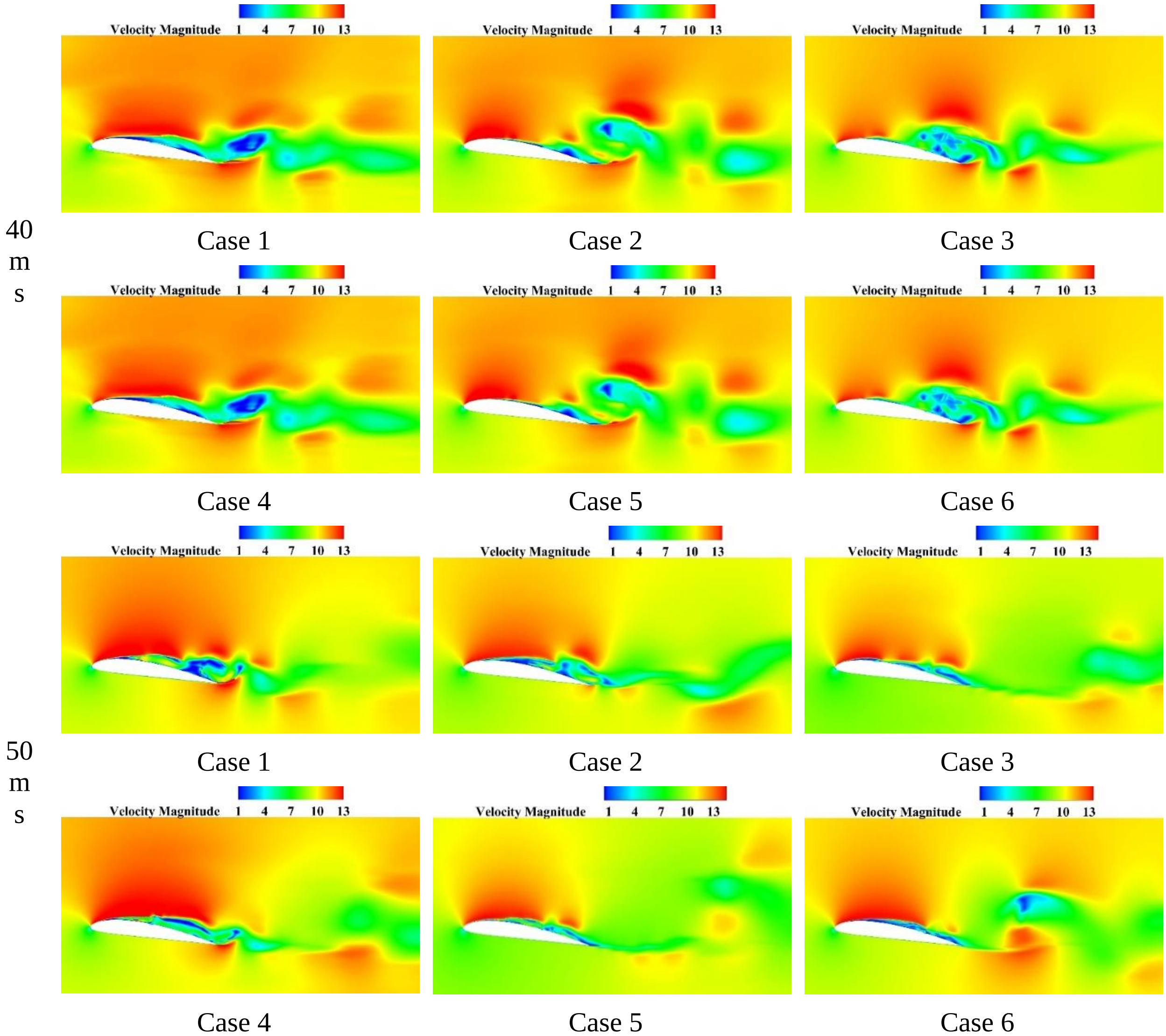

Fig. 11. Velocity-magnitude distributions for the baseline and controlled configurations at t = 30, 40, and 50 ms.

### 4.4. Pressure-field response and cycle-averaged surface pressure

Instantaneous pressure-coefficient fields are compared in Fig. 12. The baseline exhibits an extended low-pressure region over the suction side together with localized downstream pressure recovery. Changing the jet location modifies the extent and position of these low-pressure regions and the downstream recovery pattern. Forward injection affects the pressure field over a larger fraction of the suction surface, whereas rearward injection leaves more of the upstream low-pressure footprint similar to the uncontrolled case.

A more detailed inspection of Fig. 12 shows that the pressure field undergoes pronounced temporal changes and that these changes depend strongly on the chordwise injection location. At t = 30 ms, the baseline exhibits a broad low-pressure region over the suction side together with a distinct downstream pressure-recovery zone. Case 2 retains a relatively extended suction-side low-pressure region, but the pressure recovery downstream of the hydrofoil is redistributed compared with the baseline. Cases 3 and 4 also show substantial low-pressure coverage over the suction surface, accompanied by localized low-pressure structures in the downstream region, whereas Case 5 exhibits a more fragmented pressure pattern around the cavity-closure and wake regions. At t = 40 ms, the differences among the configurations become more pronounced. The baseline develops a comparatively elongated low-pressure region over the suction side, while Cases 2 and 3 exhibit a broader low-pressure footprint extending farther into the downstream flow. Case 5 also shows a pronounced low-pressure region that extends from the suction surface into a large downstream structure, whereas the pressure field of Case 6 is comparatively more localized around the hydrofoil and near wake. By t = 50 ms, the pressure distributions have reorganized again as the cavity structures evolve and are convected downstream. The baseline retains a low-pressure region over the aft suction side together with a distinct downstream structure, while Case 2 shows a clearer pressure recovery immediately downstream of the hydrofoil. Case 3 exhibits a comparatively compact low-pressure region with a weaker downstream footprint, whereas Cases 4-6 display different degrees of localized low-pressure structures and downstream recovery; in Case 6, a separated low-pressure region remains visible farther downstream. Overall, the sequence demonstrates that changing the jet location modifies not only the instantaneous extent of the suction-side low-pressure region but also the timing and spatial organization of pressure recovery in the cavity-closure and wake regions. This behavior is consistent with the location-dependent modification of cavity development and near-wall momentum redistribution observed in Figs. 9-11.

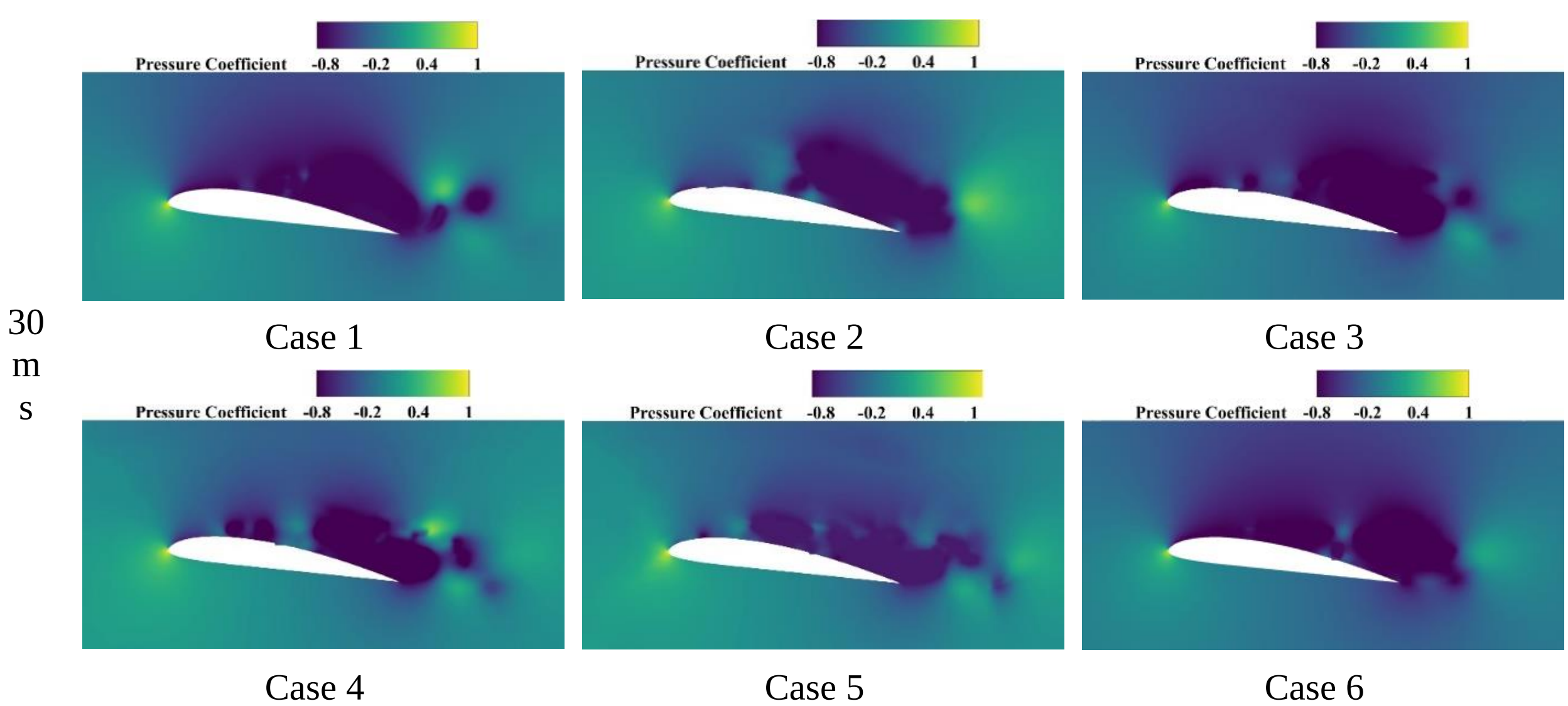

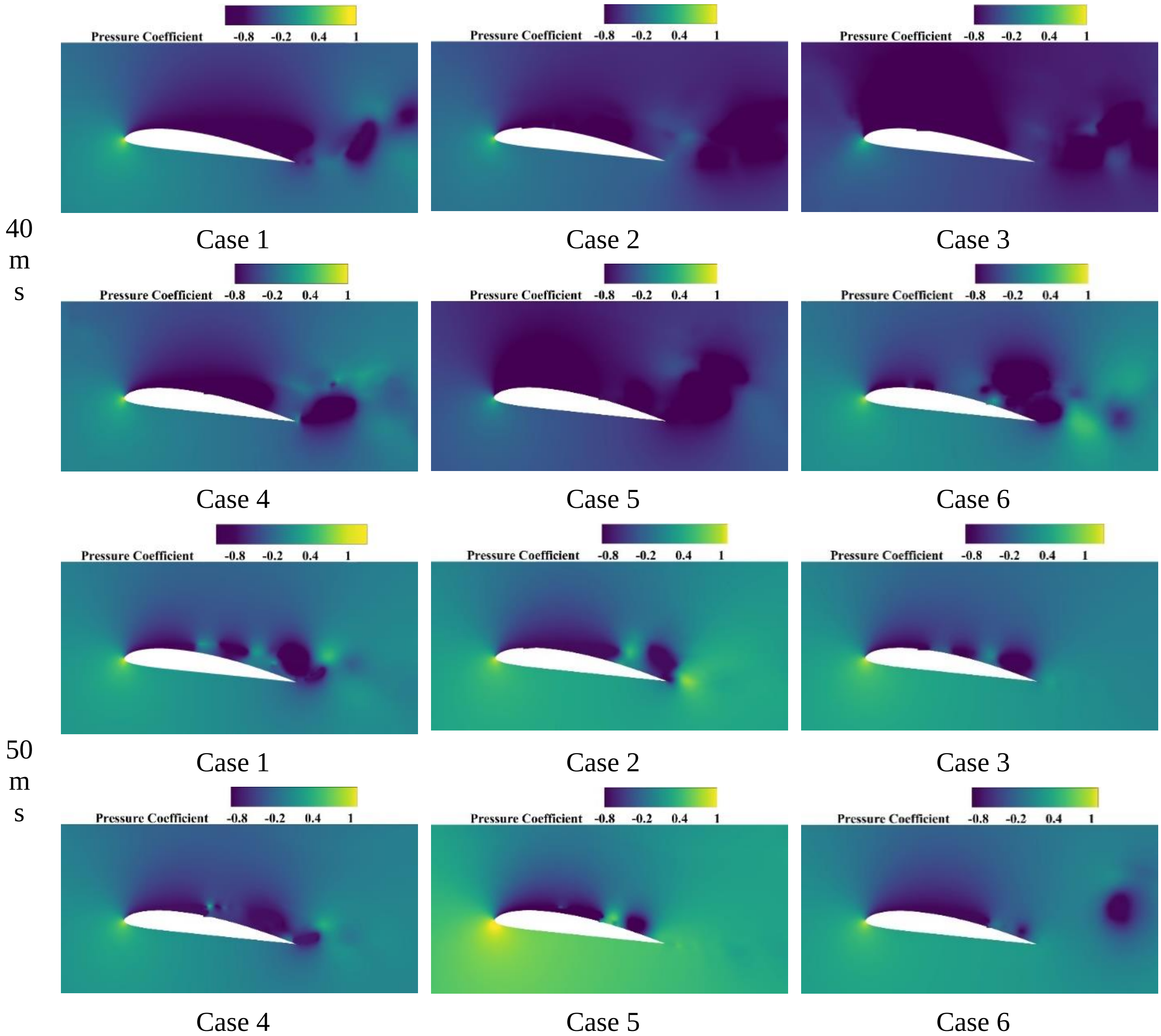

Fig. 12. Instantaneous pressure-coefficient distributions for the baseline and controlled configurations at t = 30, 40, and 50 ms.

The cycle-averaged surface distributions in Fig. 13 provide a more robust basis for comparing the six cases. For Case 2, injection at $x/c = 0.15$ alters the suction-side pressure over a comparatively long downstream distance, producing a less extensive low-pressure contribution and a smoother recovery toward the trailing edge. This redistribution is consistent with the reduction in mean lift reported later in Table 2: weakening the suction-side pressure difference reduces lift even when the drag response improves.

For the intermediate locations, the pressure curves deviate locally around and downstream of the slot while retaining a larger portion of the upstream baseline-like suction distribution. The $x/c$ = 0.30 case shows the strongest performance penalty in $C_L/C_D$, indicating that local cavity disruption and momentum addition do not necessarily translate into a favorable global force balance. The $x/c$ = 0.45 case yields a smaller drag value than the baseline, but its lift reduction largely offsets that benefit. For $x/c$ = 0.60 and 0.70, the upstream portion of the suction distribution remains comparatively close to the baseline because the imposed momentum enters after the forward cavity-development region. This progression provides a direct pressure-based explanation for why the force response is not monotonic with slot position.

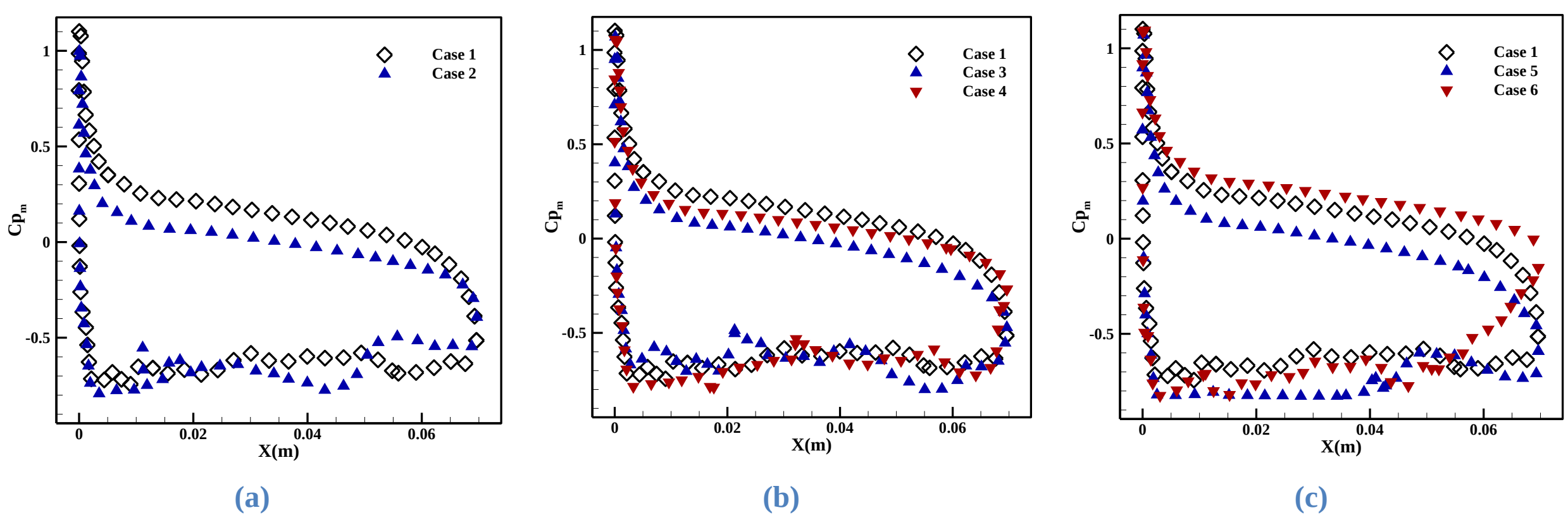

Fig. 13. Cycle-averaged surface pressure-coefficient distributions for the baseline and different chordwise injection locations: (a) Case 1 versus Case 2, (b) Case 1 versus Cases 3 and 4, and (c) Case 1 versus Cases 5 and 6.

### 4.5. Hydrodynamic performance

Table 2 summarizes the cycle-averaged lift, drag, and lift-to-drag ratio obtained from the complete second and third cavitation cycles, with the first cycle excluded from the averaging. The baseline has $C_L = 0.78$ and $C_D = 0.137$. Every controlled case reduces the cycle-averaged lift to some degree, reflecting the pressure redistribution seen in Fig. 13. The drag response is more varied: Case 2 reduces $C_D$ to 0.107, Case 4 to 0.117, and Case 6 to 0.125, whereas Cases 3 and 5 remain closer to the baseline. Thus, the benefit of injection cannot be assessed using cavity appearance alone; the force balance depends on both the pressure modification and the momentum-related losses associated with the imposed jet.

Among the investigated configurations, Case 2 (x/c = 0.15) provides the highest cycle-averaged $C_L/C_D$, equal to 6.261. Relative to the baseline, $C_D$ decreases by 21.9% and $C_L/C_D$ increases by approximately 10.0%; at the same time, $C_L$ decreases from 0.78 to 0.67. Case 2 is therefore identified specifically as the configuration with the most favorable cycle-averaged lift-to-drag ratio among the tested locations, not as a universally superior configuration. The baseline drag discrepancy of 14.1% relative to experiment (Table 1) also cautions against overinterpreting small differences between neighboring controlled cases. The Case 2-to-baseline drag change remains a clear numerical trend within the adopted framework, but its exact magnitude should be interpreted together with the validation uncertainty.

Table 2. Cycle-averaged hydrodynamic performance for the baseline and controlled configurations; the reported force coefficients are averaged over the complete second and third cavitation cycles.

| Case | $C_L$ | $C_D$ | $C_L/C_D$ |
|---|---|---|---|
| 1 | 0.78 | 0.137 | 5.693 |

| 2 | 0.67 | 0.107 | 6.261 |
|---|---|---|---|
| 3 | 0.64 | 0.127 | 5.039 |
| 4 | 0.67 | 0.117 | 5.726 |
| 5 | 0.74 | 0.131 | 5.648 |
| 6 | 0.72 | 0.125 | 5.760 |

### 4.6. Unsteady force fluctuations and spectral response

The available force histories were additionally examined to assess a second performance objective: the magnitude and spectral content of recorded unsteady hydrodynamic loading. Because the histories do not span identical cycle-resolved intervals for all configurations, this analysis is kept separate from the cycle-averaged force coefficients in Table 2. The full-record means obtained during post-processing are therefore not used to redefine the mean coefficients; instead, the RMS values are used to compare fluctuation amplitudes and the spectra are used conservatively to identify only the clearest force-response components.

The instantaneous lift- and drag-coefficient fluctuations are defined relative to the mean of each analyzed force-history record as

$$C'_L(t) = C_L(t) - \overline{C_L} \qquad (20)$$

$$C'_D(t) = C_D(t) - \overline{C_D} \qquad (21)$$

The corresponding root-mean-square fluctuation levels are defined as

$$C_{L,\mathrm{RMS}} = \sqrt{\frac{1}{T_r}\int_{t_0}^{t_0+T_r}\left[C_L(t) - \overline{C_L}\right]^2\, dt} \qquad (22)$$

$$C_{D,\mathrm{RMS}} = \sqrt{\frac{1}{T_r}\int_{t_0}^{t_0+T_r}\left[C_D(t) - \overline{C_D}\right]^2\, dt} \qquad (23)$$

where $T_r$ denotes the duration of the analyzed force-history record. These RMS values quantify the amplitude of the available recorded loading and are treated as record-based fluctuation metrics rather than statistics over a common phase-locked cavitation cycle.

The RMS comparison in Fig. 14 shows that unsteady-load stability has a different chordwise dependence from the cycle-averaged force balance. Within the available records, Case 5 (x/c = 0.60) gives the lowest fluctuation levels, with $C_{L,\mathrm{RMS}} \approx 0.112$ and $C_{D,\mathrm{RMS}} \approx 0.0165$. Relative to the baseline record, these values correspond to reductions of approximately 28.6% and 60.0%, respectively. This result contrasts directly with Case 2 (x/c = 0.15), which provides the highest cycle-averaged $C_L/C_D$ but not the lowest force-fluctuation RMS. Thus, even within the same fixed jet condition, the chordwise location favored by a mean-efficiency metric differs from that favored by the present unsteady-load metric.

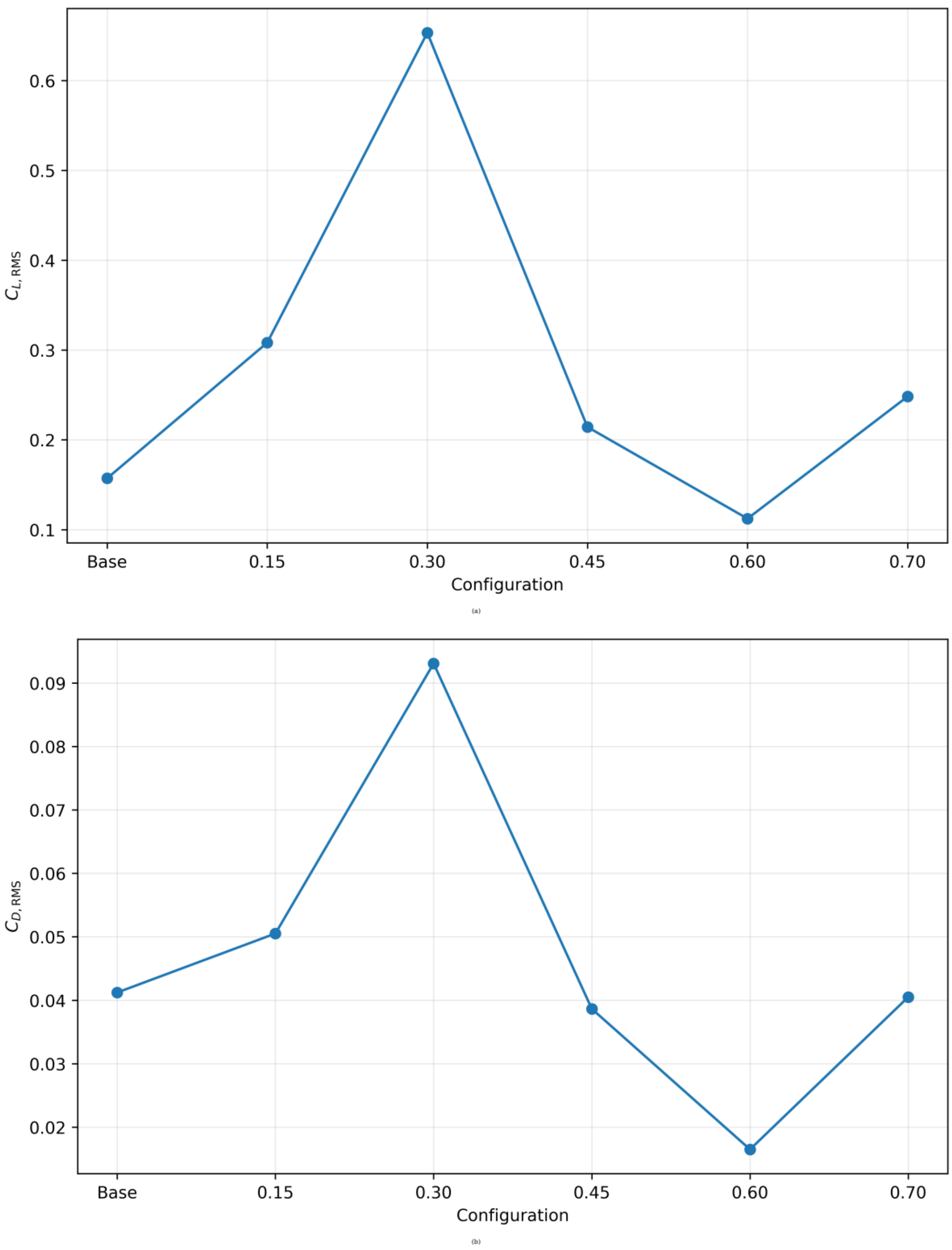

Fig. 14. RMS levels of the recorded hydrodynamic-force fluctuations for the uncontrolled baseline and five prescribed tangential micro-jet locations: (a) lift-coefficient fluctuations and (b) drag-coefficient fluctuations. The values are obtained from the available cleaned force-history records and are analyzed independently of the cycle-averaged coefficients in Table 2.

Table 3. Record-based RMS force fluctuations and robust spectral characteristics of the available force histories.

| **Case** | **Jet location x/c** | $C_{L,RMS}$ | $C_{D,RMS}$ | **Dominant force-response component** | **St_f** | **Remarks** |
|---|---|---|---|---|---|---|
| 1 | No injection | 0.157 | 0.041 | Multi-peak | — | Window-sensitive |
| 2 | 0.15 | 0.308 | 0.051 | ≈81 Hz | ≈0.57 | Clear low-frequency component in available record |
| 3 | 0.30 | 0.653* | 0.093* | Window-sensitive | — | Isolated high-amplitude excursion |
| 4 | 0.45 | 0.214 | 0.039 | Multi-peak | — | Window-sensitive |
| 5 | 0.60 | 0.112 | 0.0165 | ≈0.33 kHz (L); ≈0.17 kHz (D) | ≈2.33 (L); ≈1.16 (D) | Lowest RMS in available records; spectral components reported as force response |
| 6 | 0.70 | 0.248 | 0.041 | Window-sensitive | — | No single robust peak |

** The full-record Case 3 RMS is strongly influenced by an isolated high-amplitude load excursion and should therefore not be interpreted as evidence of persistently elevated fluctuations throughout the complete cavitation cycle.*

The Case 3 record illustrates why RMS values must be interpreted together with the underlying time history. An isolated high-amplitude load excursion substantially increases its full-record RMS, while window-sensitivity checks indicate markedly lower fluctuation levels when that event is not included in later portions of the record. The event is retained in the reported analysis, but the resulting RMS is not used to characterize Case 3 as persistently more unstable than the other configurations.

The spectral analysis was used to identify configuration-dependent force-response components. A force-based Strouhal number is defined as

$$St_f = \frac{f_f c}{U_\infty} \qquad (24)$$

where $f_f$ denotes a dominant force-response frequency, c = 0.07 m, and $U_\infty$ = 10 m/s. This definition does not establish a cavity-shedding Strouhal number because no independent cavity-volume, cavity-area, or cavity-length spectrum is available for direct frequency matching.

Fig 15 therefore focuses on the baseline and the two controlled configurations for which the clearest force-response components can be identified from the available records. Case 2 contains a low-frequency component near 81 Hz in both lift and drag, corresponding to $St_f \approx 0.57$. For Case 5, the lift spectrum contains a component near 0.33 kHz ($St_f \approx 2.33$), while the full-record drag spectrum contains a lower component near 0.17 kHz ($St_f \approx 1.16$). These frequencies are reported as force-response components only. The limited and case-dependent record durations do not support a definitive modal or harmonic interpretation, and for the remaining cases a single precise dominant frequency is not assigned because the strongest peak is sensitive to the analyzed window.

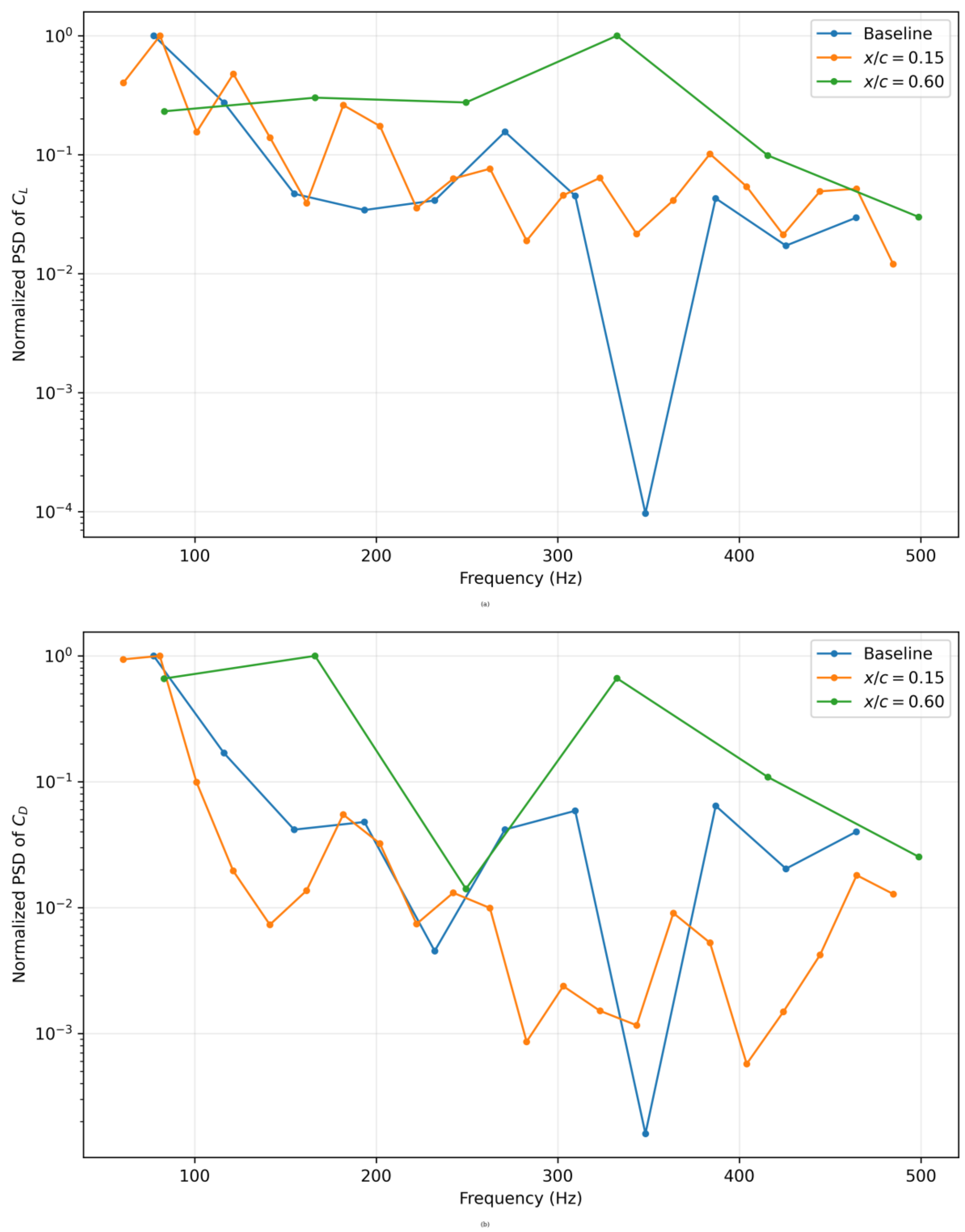

Fig. 15. Normalized force-response spectra for the uncontrolled baseline and representative controlled configurations: (a) lift-coefficient PSD and (b) drag-coefficient PSD. The spectra were obtained from linearly detrended, Hann-windowed full-record periodograms of uniformly resampled signals and are used for relative spectral comparison rather than direct validation of cavity-shedding frequency.

Taken together with Table 2, the force-history analysis establishes the principal performance contrast of the location sweep. Case 2 (x/c = 0.15) gives the highest cycle-averaged lift-to-drag ratio among the tested configurations, whereas Case 5 (x/c = 0.60) gives the lowest recorded lift- and drag-fluctuation RMS. The preferred injection location therefore depends on the selected performance objective: maximizing the present mean force ratio and minimizing the present unsteady-load metric lead to different chordwise choices. The force spectra provide supplementary information on configuration-dependent response frequencies but do not alter this RMS-based conclusion.

**4.7. Cavity-thickness dynamics**

Fig 16 compares the non-dimensional cavity thickness $d/c$ as a function of the normalized cycle coordinate $t/T$ at $x/c$ = 0.2, 0.4, and 0.6. For each configuration, $T$ is determined independently as the duration of that case's second complete cavitation-evolution cycle, from the beginning of cavity development to completion of the cycle. Consequently, $t/T$ is a case-specific cycle coordinate rather than a common absolute-time scale; equal values of $t/T$ should therefore be interpreted as corresponding normalized positions within each case's own cycle. The experimental curve and the numerical baseline are retained in each group of panels so that the controlled cases can be interpreted against the same reference. At $x/c$ = 0.2, Case 2 generally maintains a smaller thickness envelope than the baseline, whereas mid- and rear-chord injection show larger temporal variability. This is consistent with the physical expectation that a forward slot can influence the cavity early in its development before the attached vapor region has extended far downstream.

At $x/c$ = 0.4, the distinction between forward and downstream injection becomes more pronounced. The Case 2 curve remains comparatively low over much of the cycle, while the intermediate cases exhibit larger oscillations and isolated peaks. These fluctuations indicate that placing the slot within the evolving cavity region can reorganize the local interface without necessarily producing a uniformly thinner cavity. The experimental thickness exhibits larger peaks than the baseline numerical curve at several times, reinforcing the need to interpret the controlled trends as model-relative rather than as fully validated transient amplitudes.

At $x/c$ = 0.6, the curves reflect both attached-cavity evolution and passage of downstream vapor structures. Case 2 again avoids some of the larger thickness excursions seen in several downstream-injection cases, whereas Cases 3-6 display stronger case-specific variability. Taken together, the three stations show that the favorable Case 2 force response is accompanied by a generally reduced and less persistent cavity-thickness signature over the monitored chord, while moving the jet downstream shifts the control action toward local disruption rather than upstream regulation. The result supports a chordwise-coupling interpretation rather than a simple monotonic relation between jet position and instantaneous cavity size.

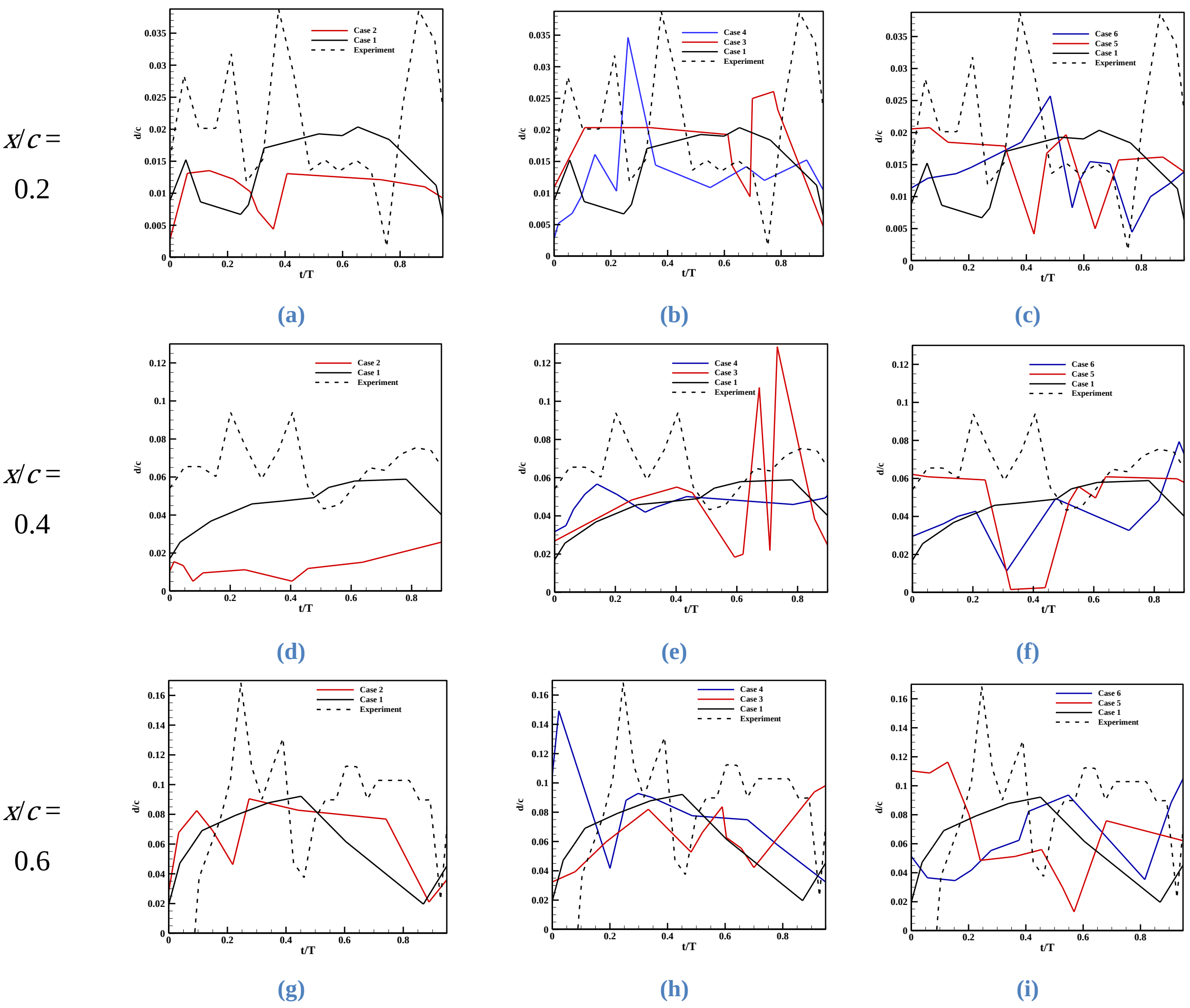

Fig. 16. Temporal variation of non-dimensional cavity thickness d/c over the case-specific normalized cycle coordinate t/T at x/c = 0.2 (a-c), x/c = 0.4 (d-f), and x/c = 0.6 (g-i). For each case, T is the independently determined duration of the second complete cavitation-evolution cycle. The uncontrolled baseline and the experimental data reported for the benchmark by Roohi et al. [6] are included for comparison.

## 4.8. Integrated physical interpretation

The combined evidence indicates that the decisive effect of slot position is the streamwise stage of cavity development at which additional tangential momentum is introduced. At $x/c = 0.15$, the jet acts while the attached cavity and near-wall momentum deficit are still developing. The resulting pressure redistribution extends over much of the downstream suction surface, large coherent vapor structures are less persistent in several sampled instants, and the mean drag decreases substantially. The associated reduction in lift is not a contradiction; it is the force-level consequence of raising the suction-side pressure while changing the cavity and wake response.

Moving the slot to $x/c = 0.30$-$0.45$ places momentum addition deeper inside the evolving cavitating region. The fields show local changes in TKE, velocity, pressure, and cavity shape, but the force balance is less favorable because upstream cavity development has already occurred and local jet-mainstream interaction can introduce additional mixing. *Re*arward placement at $x/c = 0.60$-$0.70$ leaves still more of the forward suction-side behavior close to the baseline and primarily modifies the downstream evolution. This interpretation is consistent with the broader literature showing strong configuration dependence of injection-based cavitation control [10–15] and with broader reviews showing that cavitation-control performance depends strongly on flow configuration, geometry, and operating regime [2,8,44].

The unsteady-force results add a second performance dimension to this chordwise interpretation. Although the forward location at x/c = 0.15 gives the highest cycle-averaged $C_L/C_D$, the x/c = 0.60 configuration produces the lowest record-based RMS fluctuations. The two metrics therefore identify different preferred locations under the same prescribed jet condition. This objective dependence is a central result of the present comparison: forward injection is favored when the selected criterion is the cycle-averaged lift-to-drag ratio, whereas the downstream x/c = 0.60 location is favored when the criterion is the RMS amplitude of the available force histories. The result is configuration-specific and is not advanced as a universal design rule.

The study also has three principal methodological boundaries. First, the location ranking is obtained from a two-dimensional parametric framework, whereas Fig. 8 shows that three-dimensional cavity evolution contains spanwise deformation that the planar model cannot reproduce. Second, the jet is prescribed and therefore does not include the hydraulic coupling, energetic cost, or self-regulation of a pressure-fed internal channel. Third, baseline validation is strongest for cycle-averaged pressure and lift; the larger drag discrepancy and the limited transient experimental comparison constrain the certainty of absolute force and phase-resolved quantities. These boundaries define the scope of the conclusions while preserving the value of the internally consistent case-to-case comparison.

## 5. Conclusions

1. Under the investigated $Re = 7 \times 10^5$ and $\sigma = 0.8$ condition, changing the chordwise location of a prescribed tangential micro-jet substantially modifies the coupled cavity, near-wall momentum, pressure, and wake response. The effect is therefore not a simple monotonic function of downstream position.

2. Forward injection at x/c = 0.15 provides the highest cycle-averaged lift-to-drag ratio among the tested locations. Relative to the baseline, $C_D$ decreases from 0.137 to 0.107 (21.9%) and $C_L/C_D$ increases from 5.693 to 6.261 (approximately 10.0%), while $C_L$ decreases from 0.78 to 0.67. The improvement is thus specific to the lift-to-drag metric and should not be interpreted as a simultaneous increase in lift.

3. The vapor, velocity, normalized-TKE, pressure, and cavity-thickness results indicate that forward injection acts while the attached cavity and near-wall momentum deficit are still developing. Moving the slot downstream introduces momentum after a larger upstream cavity and momentum deficit have already formed, shifting the control action from earlier streamwise regulation toward more localized downstream modification.

4. Unsteady-load stability identifies a different preferred location. Case 5 ($x/c = 0.60$) yields the lowest recorded $C_L$ and $C_D$ fluctuation RMS values, approximately 0.112 and 0.0165, corresponding to reductions of about 28.6% and 60.0% relative to the baseline record. Hence, the location that maximizes the present cycle-averaged $C_L/C_D$ does not coincide with the location that minimizes the present recorded force fluctuations.

5. The cycle-averaged pressure distributions and cavity-thickness histories support this objective-dependent interpretation. Case 2 modifies the suction-side pressure over a comparatively broad downstream region and generally reduces the monitored cavity-thickness signature, whereas intermediate and rearward locations retain more upstream baseline-like behavior and produce stronger case-specific local variability.

6. The conclusions are bounded by the numerical framework. The five-location ranking is based on two-dimensional screening; the representative three-dimensional Case 3 calculation illustrates spanwise cavity deformation but does not validate a three-dimensional ranking. The jet is prescribed rather than pressure-driven, and the transient experimental validation does not extend to force spectra or directly measured cavity-shedding frequency.

Future work should test whether the objective-dependent location preference persists across additional cavitation numbers and jet momentum ratios, extend the comparison to broader three-dimensional case coverage, and couple the external cavitating flow to a pressure-driven internal-channel model where passive or self-regulated operation is of interest. Direct spectra of cavity volume, cavity area, vapor-volume integral, or cavity length would also permit a defensible comparison between cavity-shedding dynamics and the force-response frequencies identified here.

## Data availability

The data supporting the findings of this study are available from the corresponding author upon reasonable request.

## Declaration of competing interest

The authors declare that they have no known competing financial interests or personal relationships that could have appeared to influence the work reported in this paper.